%% file: main.tex
\documentclass[%
 reprint,
nofootinbib,
 amsmath,amssymb,
 aps,
]{revtex4-2}

\usepackage{graphicx}
\usepackage{amsmath,amssymb}
\usepackage{braket}
\usepackage{xcolor}
\usepackage{soul}
\usepackage[colorlinks=true,
            linkcolor=black,
            citecolor=black,
            urlcolor=blue]{hyperref}
\usepackage[nameinlink]{cleveref}
\usepackage{slashed}
\usepackage{placeins}
\usepackage{bm}
\usepackage{bbm}
\definecolor{turquoise}{RGB}{64,224,208}
\colorlet{MODCOLOR}{black}
\colorlet{MODCOLOR2}{black}
\colorlet{MODCOLOR3}{black}
\input{include_2cols}

\begin{document}

\title{Extraction of spectral densities from lattice correlators: decoupling signal from noise}

\author{Alessandro Lupo}
\email{alessandro.lupo@cpt.univ-mrs.fr}
\affiliation{Aix Marseille Univ, Université de Toulon, CNRS, CPT, Marseille, France}

\author{Nazario Tantalo}
\email{nazario.tantalo@roma2.infn.it}
\affiliation{Dipartimento di Fisica, Università di Roma Tor Vergata and INFN, Sezione di Roma Tor Vergata,
    Via della Ricerca Scientifica 1, I-00133 Rome, Italy}

\setlength\abovedisplayskip{10pt}
\setlength\belowdisplayskip{10pt}

\setlength{\parskip}{14pt}
\setlength{\parindent}{0pt}

\begin{abstract}
We expand the treatment of the problem of the extraction of smeared spectral densities from Euclidean correlators, introduced in Ref.~\cite{Hansen:2019idp}, providing an alternative which does not rely on the Backus-Gilbert regularization.
This is possible due to the observation that the solution can be decomposed into a sum of terms, in the spirit of the singular value decomposition, where those with the largest contribution to the statistical noise happen to contribute the least to the central value of the smeared spectral density.
The analysis of the systematics of the inverse problem is then shifted to finding the optimal truncation of such summation, so that the signal is saturated before the noise explodes.
We scrutinise the performance and systematics of this approach either as a standalone procedure, or to complement the stability analysis required to extrapolate the unbiased result in the Backus-Gilbert regulated version of the solution.
\end{abstract}

\maketitle

\input{sections/introduction}
\input{sections/theoreticalframework}
\input{sections/newmethod}
\input{sections/pulls}
\input{sections/conclusions}
\section*{Acknowledgements}
A.L. is funded in part by l’Agence Nationale de la Recherche (ANR), under grant ANR-22-
CE31-0011. We thank the ETM collaboration for providing us the data for the statistical correlation matrix used in this study. We warmly thank A.~De Santis and F.~Margari for their help and for very useful discussions. 
\appendix
\input{sections/appendix}

\newpage
\bibliographystyle{JHEP}
\bibliography{main}

\end{document}

%% file: include_2cols.tex
\usepackage{bm}
\usepackage{bbm}

\newcommand{\LO}{\ifmmode{\mathrm{LO}}\else{LO}\fi}
\newcommand{\NLO}{\ifmmode{\mathrm{NLO}}\else{NLO}\fi}
\newcommand{\NNLO}{\ifmmode{\mathrm{NNLO}}\else{NNLO}\fi}

\newcommand{\fm}{\ifmmode{\texttt{fm}}\else{\texttt{fm} }\fi}
\newcommand{\Op}{\ifmmode{\mathcal{O}}\else{$\mathcal{O}$}\fi}
\newcommand{\Cov}{\ifmmode{\mathrm{Cov}}\else{Cov}\fi}

\renewcommand{\vec}[1]{\bm{#1}}
\renewcommand{\matrix}[1]{\mathbbmss{#1}}

\newcommand{\globalWidth}{0.9\columnwidth}

%% file: sections/introduction.tex
\section{Introduction}

{\color{MODCOLOR} The operation of} extracting spectral densities from correlation functions in Euclidean time is of primary importance in particle physics. Euclidean correlation functions are in fact accessible through systematically improvable lattice simulations, which are used to probe non-perturbatively strongly interacting gauge theories such as Quantum Chromo Dynamics (QCD). While the lattice can provide Euclidean correlators, the real-time Minkowskian dynamics, required to make predictions for scattering experiments, is encoded in the associated spectral functions. In practice, their extraction from Euclidean correlators amounts to inverting a Laplace transform on a finite set of noisy data. The difficulties of this operation pose a limit to the scope of lattice simulations, and progress in our theoretical understanding of QCD and other strongly-interacting theories.

Given its importance, the topic has been long studied. Already in Ref.~\cite{Barata:1990rn} the problem was formally solved by showing that scattering amplitudes can be approximated arbitrarily well by appropriate combinations of Euclidean correlators. To achieve a practical solution, though, one needs to deal with the major challenge of the statistical noise. The mathematics required to numerically address this class of problems is rather old and well established, see e.g.~\cite{Bertero_1988}, and it is generally understood that making progress in this field requires devising algorithmic solutions which provide robust {\color{MODCOLOR} numerical} estimates of the errors.

Several approaches have been proposed to cope with the spectral densities inverse problem, yet phenomenological applications remained limited due to the difficulties in understanding and controlling systematic effects, or due to model dependence. In Ref.~\cite{Hansen:2019idp}, together with Martin Hansen, we proposed a model-independent approach to solve the inverse problem, elements of which can be found in previous literature~\cite{Furmanski:1981ja, 10.1111/j.1365-246X.1968.tb00216.x, pijpers1994sola}, {\color{MODCOLOR} focusing on the possibility to provide reliable numerical estimates of the errors}. In this approach, the main systematic is inherited by the Backus-Gilbert~\cite{10.1111/j.1365-246X.1968.tb00216.x} regulator, parameterised by a trade-off real parameter $\lambda>0$, which biases the solution. The {\color{MODCOLOR} numerical} extrapolation to the unbiased result at $\lambda=0$ is performed through a stability analysis~\cite{Bulava:2021fre}. Such procedure has been since then placed under severe scrutiny, using Monte-Carlo lattice simulations in the controlled environment of the two dimensional $O(3)$ non-linear $\sigma$-model~\cite{Bulava:2021fre}, with closure tests for realistic four dimensional QCD simulations~\cite{DelDebbio:2024lwm, Buzzicotti:2023qdv}, and with comparison with other spectroscopy techniques~\cite{DelDebbio:2022qgu, Bennett:2024cqv}. The encouraging results obtained in these contexts \textcolor{MODCOLOR}{provided empirical evidence that the numerical procedure adopted to remove the regulator is effective}, and motivated the beginning of studies with phenomenological relevance and granted the first ab-initio calculations of certain inclusive decay rates and cross-sections~\cite{ExtendedTwistedMassCollaborationETMC:2022sta, ExtendedTwistedMass:2024myu, Evangelista:2023fmt, DeSantis:2025qbb, DeSantis:2025yfm, Bonanno:2023ljc, Bonanno:2023thi}.

In this work we propose a different way to regulate the reconstruction proposed in Ref.~\cite{Hansen:2019idp}, which allows one to work at any value of $\lambda$, including zero. This is achieved by noticing that the smeared spectral function, written as a linear combination of the input correlation functions, can be decomposed into a basis such that the terms contributing the most to the noise do not contribute to the signal. Truncating the sum in this basis therefore provides a well-behaved result without the need for a Backus-Gilbert regulator, which remains however possible. We will describe the systematics of such truncation, show its performance against, and in conjunction with, the Backus-Gilbert regulator. The goal is to improve on the reliability of the results obtained with the stability analysis, which will culminate, in this work, in the proposal of a hybrid procedure. Given the simplicity of the new regularisation, we also envision its standalone use as an efficient way to get a quick reconstruction of the smeared spectral density. However, as we will show, the procedure on its own tends to be either overly conservative, or too aggressive, depending on where the sum is truncated, which further motivates the hybrid approach.

For completeness, we stress that many other methods have been proposed, yet to our knowledge only a few of them have been used in phenomenological studies. The original Backus-Gilbert proposal and Bayesian methods have been popular for a while~\cite{Burnier:2013nla, Hansen:2017mnd}, but the unavoidable systematics related to the smearing of the spectral density are hard to address within these approaches. Nonetheless, in the context of parton distribution functions, Bayesian methods have been shown to produce results which are consistent, within available precision, with the initial data~\cite{DelDebbio:2020rgv, DelDebbio:2021whr, Medrano:2025cmg}. Moreover, a version of the Bayesian setup which takes explicitly into account the smearing was given in Ref.~\cite{DelDebbio:2024lwm} and was shown to be partially analogous to Ref.~\cite{Hansen:2019idp}. \textcolor{MODCOLOR}{Chebyshev} polynomials have also been recently used to provide inclusive decays with {\color{MODCOLOR} reliable numerical estimates of the errors}~\cite{Barata:1990rn, Barone:2023tbl, Kellermann:2025pzt}. Finally, we mention alternative approaches which have been recently proposed, such as Refs.~\cite{Bergamaschi:2023xzx, Bruno:2024fqc} that offer different  perspectives to attack the problem,~\cite{Buividovich:2025hmn} for applications at high temperature, {\color{MODCOLOR}and Ref.~\cite{Tsuji:2026zku} which follows an approach similar to the one of this paper and that appeared on-line shortly after this work}.

In Section~\ref{sec:theo_frame} we introduce our notation for the inverse problem and its solution in the exact case (i.e. without statistical errors). For a solution to be useful when errors are present, a regulator must be introduced, together with a reliable procedure to remove it within the given precision. Two of such regulators, and the {\color{MODCOLOR} numerical} procedures to {\color{MODCOLOR} estimate their effects}, are introduced in Section~\ref{sec:regularisations}: first the stability analysis to remove the Backus-Gilbert regulator (\ref{sec:stability}), then a new approach is described (\ref{sec:decoupling_signal_from_noise}). We systematically test the performance of the new procedure in Section~\ref{sec:pulls} by performing closure tests using statistical errors from state-of-the-art lattice simulations. These tests allowed us to tune the algorithm and design a solid {\color{MODCOLOR} numerical} procedure which merges aspects of our previous work, Ref.~\cite{Hansen:2019idp}, with our new findings. We draw our conclusions in Section~\ref{sec:conclusions} and discuss in appendix~\ref{sec:appendix} the interplay of the {\color{MODCOLOR} numerical} extrapolations which are required to remove the lattice and the algorithmic regulators.

%% file: sections/theoreticalframework.tex
\section{Theoretical Framework}\label{sec:theo_frame}

In this section we retrace the ideas behind the starting point of this work, the procedure described in Ref.~\cite{Hansen:2019idp}, often dubbed the HLT method, and set the stage for the introduction of the new regularization procedure.

We are concerned with the extraction of smeared spectral densities from Euclidean correlators. As in Ref.~\cite{Hansen:2019idp}, to simplify the discussion, we consider below the simplest case, i.e.\ a spectral density  $\rho(E)$ depending upon a single energy variable $E$ and the associated correlation function $C(t)$ of two gauge-invariant operators separated by a Euclidean time $t$. The generalization of the formalism and of the HLT method to the case of multivariate spectral densities is thoroughly discussed in Ref.~\cite{Patella:2024cto}.

We consider the situation in which the correlator is known at a finite number of points $t=n\tau$, where $\tau$ is a length scale in physical units and $n=1\dots N$ is an integer. 
The relation between the spectral density and the correlator is then
\begin{flalign}
    C(n \tau) = \int_{E_0}^\infty dE \; b(n\tau,E) \,\rho(E) \, ,
    \label{eq:laplace_transform}
\end{flalign}
where $E_0\ge 0$, \textcolor{MODCOLOR}{and in the simplest case},
\begin{flalign}
    b(n\tau,E) = e^{-n \tau E} \, .
    \label{eq:simplebasis}
\end{flalign}

In quantum field theories, spectral densities are tempered distributions and therefore, in order to cope with them in numerical applications, the introduction of a smearing kernel is unavoidable. The infinite amount of physical information contained into $\rho$ as a function of $E$ can then be extracted by studying the functional
\begin{flalign}
    \rho[\mathcal{S}] = \int_{E_0}^{\infty} dE \, \mathcal{S}(E) \, \rho(E) \, ,
    \label{eq:smeared_rho_def}
\end{flalign}
the smeared spectral density, by choosing different smearing kernels $\mathcal{S}(E)$. Notice that the correlator $C(n\tau)$, defined in Eq.~\eqref{eq:laplace_transform},  is itself a smeared spectral density, i.e.\ $C(n\tau)=\rho[b(n\tau)]$.  

The smearing kernel belongs to the space $\mathcal{L}_2[E_0,\infty]_\alpha$, i.e.\ to the set of real functions $f$ such that
\begin{flalign}
\left\| f \right\|_\alpha^2 = \int_{E_0}^{\infty} dE \, e^{\alpha \tau E }\,  \left[f(E)\right]^2 < \infty \,,
\label{eq:norm_def}
\end{flalign}
where, for later use, we conveniently introduced the real parameter $\alpha <2$. In the numerical part of this work, we will always use $E_0=0$. If the set of functions that can be represented as linear combinations of the basis functions is dense in $\mathcal{L}_2[E_0,\infty]_\alpha$,  the kernel\footnote{Since in general $\rho(E)$ is a tempered distribution $\mathcal{S}(E)$ needs to be a Schwartz function for $\rho[\mathcal{S}]$ to be a finite real number. However, the approximation methodology we are describing applies, {\color{MODCOLOR} by setting $\alpha=0$}, to the larger set of square integrable functions.} $\mathcal{S}(E)$ can \emph{exactly} be represented as
\begin{flalign}
&
\mathcal{S}_{\scriptscriptstyle N}(E) = \sum_{n=1}^N g_{\scriptscriptstyle [\mathcal{S}],N}(n)\, b(n\tau,E)\,,
\nonumber \\[8pt]
&
\mathcal{S}(E) = \lim_{N\mapsto \infty} \mathcal{S}_{\scriptscriptstyle N}(E)\;.
\label{eq:exact_representation_of_kernel}
\end{flalign}
This is certainly true for the functions given in Eq.~\eqref{eq:simplebasis} for \emph{any} choice of $\tau>0$. In fact, by considering the change of variables $x=\exp(-\tau E)$, one has $b(n\tau,E)=x^n$  and the problem is mapped to that of the polynomial representation of the kernel $\mathcal{S}(-\log(x)/\tau)$ in $x\in (0,1]$. Eq.~\eqref{eq:exact_representation_of_kernel} is then a direct consequence of the classical Weierstrass theorem (see appendix E.2 of Ref.~\cite{Patella:2024cto} for more details). In practice we must work at finite $N$, but the theorem guarantees that the approximation can become arbitrarily accurate by increasing $N$.
This holds true either if $\tau$ is kept fixed in physical units or if it is identified with the lattice spacing (see the appendix~\ref{sec:appendix} for more details concerning this point).

Denote $\vec{g} \in \mathbb{R}^N$ as the vector of elements $g(n)$. Following again Ref.~\cite{Hansen:2019idp}, the coefficients in Eq.~\eqref{eq:exact_representation_of_kernel} are defined by
\begin{flalign}
    \vec{g}_{\scriptscriptstyle [\mathcal{S}],N} = \underset{\vec{g} \in \mathbb{R}^N}{\mathrm{argmin}} \; A[\vec{g}] \, ,
    \label{eq:g_as_argmin_of_A}
\end{flalign}
with
\begin{flalign}
    A[\vec{g}] =\left\| \mathcal{S}(E)-\sum_{n=1}^N g(n) b(n\tau,E)  \right\|_{\alpha}^2 \,.
\label{eq:functional_A}
\end{flalign}
The smeared spectral density of Eq.~\eqref{eq:smeared_rho_def} can then be written as
\begin{flalign}
&
 \rho_{\scriptstyle N}[\mathcal{S}] = 
 \sum_{n=1}^N C(n \tau)\ g_{\scriptscriptstyle [\mathcal{S}],N}(n) 
\equiv
 \vec{C}_{\scriptstyle N}^T \cdot \vec{g}_{\scriptscriptstyle [\mathcal{S}],N} 
 \, ,
 \label{eq:rho_as_sum_g_dot_c}
 \nonumber \\[8pt]
&
\rho[\mathcal{S}]=\lim_{N\mapsto \infty} \rho_{\scriptstyle N}[\mathcal{S}]\, , 
\end{flalign}
where $\vec{C}_{\scriptstyle N} \in \mathbb{R}^N$ is a vector collecting the values of the correlator.

Eq.~\eqref{eq:g_as_argmin_of_A} is equivalent to a linear system of equations, whose solution is given by
\begin{flalign}
    \vec{g}_{\scriptscriptstyle [\mathcal{S}],N} = \matrix{A}_{\scriptstyle N}^{-1}\cdot \vec{f}_{\scriptscriptstyle [\mathcal{S}],N}\,,
    \label{eq:g_equals_A_dot_f}
\end{flalign}
where the $N \times N$ matrix $\matrix{A}_{\scriptstyle N}$ and the $N$-dimensional kernel vector $\vec{f}_{\scriptscriptstyle [\mathcal{S}],N}$ have elements
\begin{flalign}
&
    A(n,m) = \int_{E_0}^\infty dE \, e^{\alpha\tau E}\, b(n\tau,E) \, b(m\tau, E) \, ,
\nonumber \\[8pt]    
&
f_{\scriptscriptstyle [\mathcal{S}]}(n)=    \int_{E_0}^\infty dE \, e^{\alpha\tau E}\, b(n\tau,E) \, \mathcal{S}(E) \, .
\end{flalign}
By plugging Eq.~\eqref{eq:g_equals_A_dot_f} into Eq.~\eqref{eq:rho_as_sum_g_dot_c} we thus get
\begin{flalign}
    \rho_{\scriptstyle N}[\mathcal{S}] 
    = \vec{C}_{\scriptstyle N}^T \cdot \vec{g}_{\scriptscriptstyle [\mathcal{S}],N}
    = \vec{C}_{\scriptstyle N}^T \cdot \matrix{A}_{\scriptstyle N}^{-1}\cdot \vec{f}_{\scriptscriptstyle [\mathcal{S}],N}\,.
    \label{eq:rho_equals_C_dot_A_dot_f}
\end{flalign}

\begin{figure}[tb!]
    \centering
    \includegraphics[width=\globalWidth]{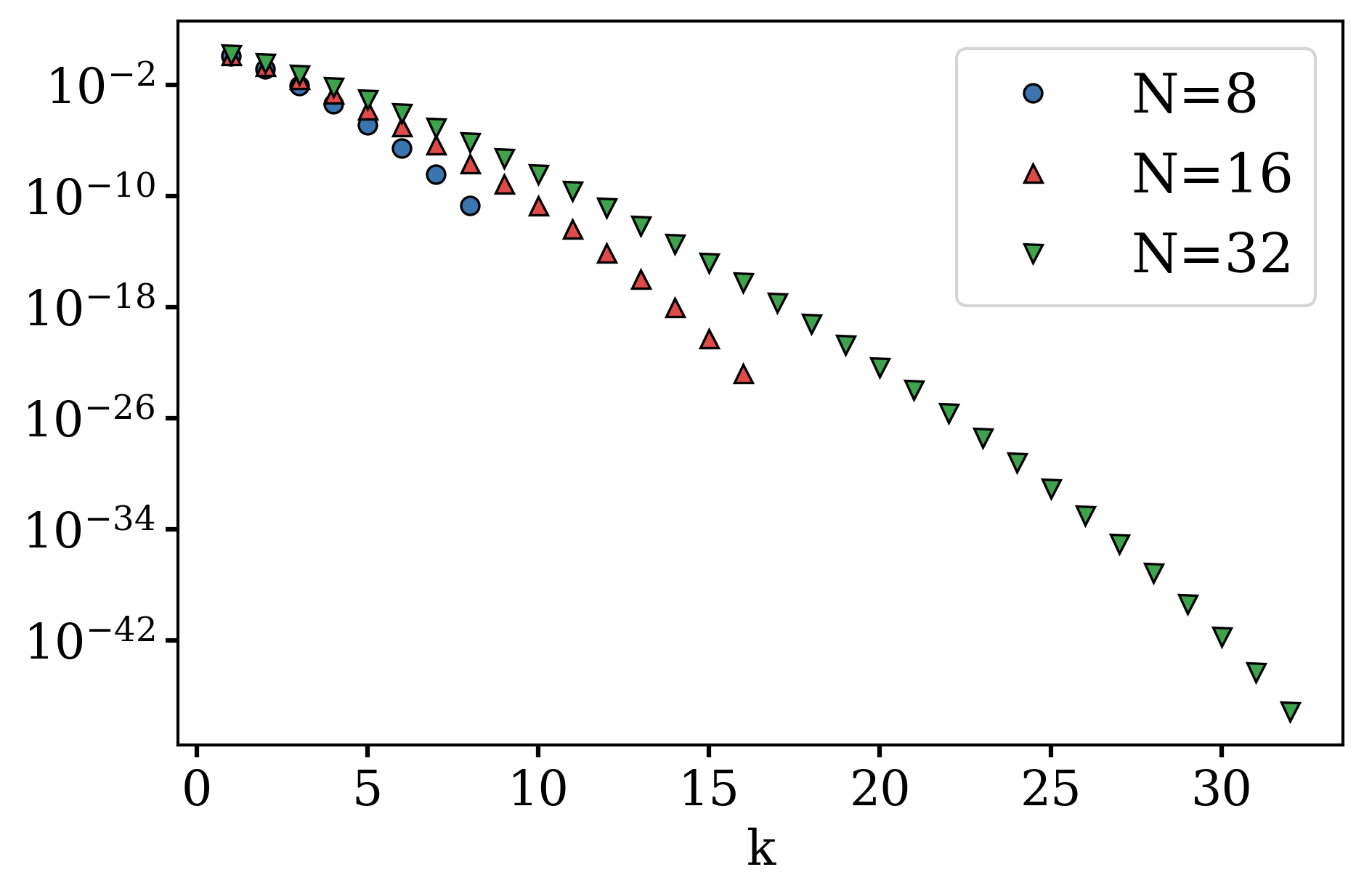}
    \caption{Eigenvalues $a_N(k)$ of the matrix $\matrix{A}_{\scriptstyle N}$ for different values of $N$.}
    \label{fig:eigenvalues_of_A}
\end{figure}

The real and symmetric matrix $\matrix{A}_{\scriptstyle N}$ is a special case of a Cauchy matrix, whose condition number grows exponentially with $N$, as shown in Fig.~\ref{fig:eigenvalues_of_A} for different values of $N$.
As we will show, such ill-conditioning results in the impossibility to handle correlation functions with any realistic level of statistical noise. In order to extract  $\rho[\mathcal{S}]$ from $\vec C_{\scriptstyle N}$ it is necessary to introduce an intermediate regulator \emph{and} a {\color{MODCOLOR} numerical procedure to remove it, for instance by extrapolating it to zero, with reliable estimates of the associated errors}.

To better illustrate this crucial point, and to set the stage for the introduction of the new approach proposed in this work, we now rotate the previous formulae, derived in terms of Euclidean time and therefore expressed in terms of time-vectors, to the basis in which the matrix $\matrix{A}_{\scriptstyle N}$ is diagonal.

Let $a_{\scriptstyle N}(k)$ be the eigenvalues of $\matrix{A}_{\scriptstyle N}$, which we order according to
\begin{flalign}
a_{\scriptstyle N}(1)  >  a_{\scriptstyle N}(2)  > \cdots > a_{\scriptstyle N}(N) \,,
\end{flalign}
and $\vec{u}_{\scriptstyle N}(k)$ the corresponding eigenvectors,
\begin{flalign}\label{eq:Au_au}
\matrix{A}_{\scriptstyle N}\cdot \vec{u}_{\scriptstyle N}(k) = a_{\scriptstyle N}(k)\, \vec{u}_{\scriptstyle N}(k)\, .
\end{flalign}
The eigenvectors can be chosen to satisfy
\begin{flalign}
\vec{u}^T_{\scriptstyle N}(k)\cdot \vec{u}_{\scriptstyle N}(h) = \delta_{\scriptstyle k,h}\;,
\qquad
\sum_{k=1}^N \vec{u}_{\scriptstyle N}(k)\vec{u}^T_{\scriptstyle N}(k) = \matrix{1}_{\scriptstyle N}\;.
\end{flalign}
For any time-vector $\vec{v}_{\scriptstyle N}$, living in time-space and having elements $v_{\scriptstyle N}(n)$, we now introduce {\color{MODCOLOR} its representation $\vec{\hat v}$ in the orthonormal eigenbasis $\{u_N(k)\}$ of $\matrix{A}_N$ (the eigen-space) with elements}
\begin{flalign}
\hat{v}_N(k) = \vec{u}^T_{\scriptstyle N}(k)\cdot \vec{v}\;.
\end{flalign}
By using this notation, the {\color{MODCOLOR} time-vectors $\vec{C}_{\scriptstyle N}$, $\vec{f}_{\scriptstyle [\mathcal{S}],N}$ and $\vec{g}_{\scriptstyle [\mathcal{S}],N}$
are, in the new basis,} $\vec{\hat{C}}_{\scriptstyle N}$,  $\vec{\hat{f}}_{\scriptstyle [\mathcal{S}],N}$ and $\vec{\hat{g}}_{\scriptstyle [\mathcal{S}],N}$ and have components
\begin{flalign}
&
\hat{C}_{\scriptstyle N}(k)= \vec{u}^T_{\scriptstyle N}(k)\cdot \vec{C}_{\scriptstyle N}\,,
\nonumber \\[8pt]
&
\hat{f}_{\scriptstyle [\mathcal{S}],N}(k)= \vec{u}^T_{\scriptstyle N}(k)\cdot \vec{f}_{\scriptstyle [\mathcal{S}],N}\,,
\nonumber \\[8pt]
&
\hat{g}_{\scriptstyle [\mathcal{S}],N}(k)= \vec{u}^T_{\scriptstyle N}(k)\cdot \vec{g}_{\scriptstyle [\mathcal{S}],N}
= \frac{\hat{f}_{\scriptstyle [\mathcal{S}],N}(k)}{a_{\scriptstyle N}(k)}\;.
\end{flalign}
This allows one to write the following representation of $\rho_{\scriptstyle N}[\mathcal{S}]$ in eigen-space,
\begin{flalign}
    \rho_{\scriptstyle N}[\mathcal{S}] 
    = \vec{\hat{C}}_{\scriptstyle N}^T \cdot \vec{\hat{g}}_{\scriptscriptstyle [\mathcal{S}],N}
    = \sum_{k=1}^N \frac{\hat{C}_{\scriptstyle N}(k)\, \hat{f}_{\scriptstyle [\mathcal{S}],N}(k)}{a_{\scriptstyle N}(k)}\,,
    \label{eq:rho_dual}
\end{flalign}
which is equivalent to the time-space representation given in Eq.~\eqref{eq:rho_equals_C_dot_A_dot_f}. Such diagonal representation appears already in Ref.~\cite{Bertero_1988}, and was {\color{MODCOLOR} profitably used in the more recent} Ref.~\cite{Bruno:2024fqc}.

The consequences of the ill-conditioning of $\matrix{A}_{\scriptstyle N}$ can be now made explicit. Denoting the statistical fluctuations of the correlator as
\begin{flalign}
\Delta_\mathrm{stat}\vec{ C}_{\scriptstyle N} = \vec C_{\scriptstyle N} - \langle \vec C_{\scriptstyle N} \rangle \, ,
\end{flalign}
the statistical fluctuations induced on the smeared spectral density are given by
\begin{flalign}
    \Delta_\mathrm{stat} \rho_{\scriptstyle N}[\mathcal{S}] 
    = \sum_{k=1}^N \frac{\Delta \hat{C}_{\scriptstyle N}(k)\, \hat{f}_{\scriptstyle [\mathcal{S}],N}(k)}{a_{\scriptstyle N}(k)}\,.
\end{flalign}
\begin{figure}[t!]
    \centering
    \includegraphics[width=\globalWidth]{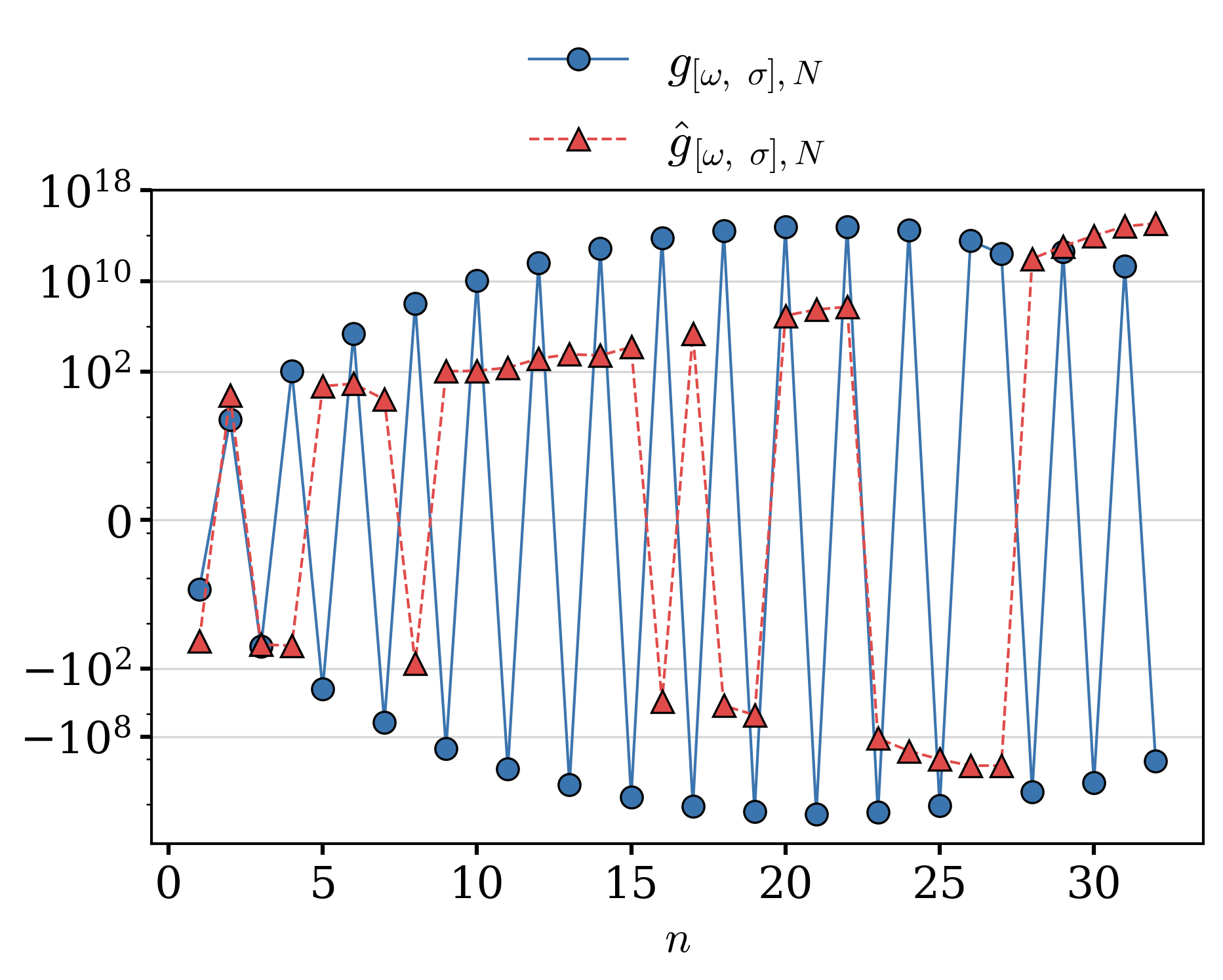}
    \caption{Values of the coefficients used to approximate a Gaussian smearing kernel, with $N=32$, $\tau\omega=0.5$ and $\tau\sigma=0.5$ (see Eqs.~\eqref{eq:gaussian} and~\eqref{eq:rhogaussian}). The time-space coefficients are shown in blue while the corresponding eigen-space coefficients are shown in red. In the eigen-space, the magnitude of the coefficients changes at a much slower rate and almost monotonically, which will be relevant for our new regularisation in Section~\ref{sec:decoupling_signal_from_noise}. To see how the values of these coefficients depend on the smearing kernel, see Fig.~\ref{fig:g_change_omega_sigma}.}
    \label{fig:g_shared}
\end{figure}
Unless $\hat{f}_{\scriptstyle [\mathcal{S}],N}(k)$ decays fast enough to suppress the increasingly large contributions from the eigenvalues $a_N^{-1}(k)$ (see Fig.~\ref{fig:eigenvalues_of_A}), the time-space coefficients will also grow exponentially in magnitude
\begin{flalign}\label{eq:g_k_basis}
    \vec{g}_{\scriptscriptstyle [\mathcal{S}],N} = 
    \sum_{k=1}^N
    \vec{u}_{\scriptstyle N}(k)\,
    \frac{\hat{f}_{\scriptstyle [\mathcal{S}],N}(k)}{a_{\scriptstyle N}(k)}\,.
\end{flalign}
We now wish to show typical behaviour of these coefficients in both the time-space and the eigen-space. It will be common, throughout this work, to choose as smearing kernel a Gaussian, and probe different energy regions by changing its centre and/or width. In this case we have
\begin{flalign}
&
\mathcal{S}(E) \quad \rightarrow \quad \frac{e^{-\frac{(E-\omega)^2}{2\sigma^2}}}{\sqrt{2\pi}\sigma}\;,
\label{eq:gaussian}
\end{flalign}
and we use the notation
\begin{flalign}
\rho[\mathcal{S}] \quad \rightarrow \quad \rho[\omega,\sigma] \;.
\label{eq:rhogaussian}
\end{flalign}
With this choice, and fixing $\alpha=0$ for the remainder of this section, we show examples of the coefficients in Fig.~\ref{fig:g_shared} for illustrative values of the Gaussian parameters, $\tau\omega=0.5$ and $\tau\sigma=0.5$, in both the time (blue) and in the eigen-space (red). {\color{MODCOLOR2} The size $N=32$ is chosen because, for the cases considered below, it suffices to perform a reliable  $N\rightarrow \infty$ numerical extrapolation by checking for convergence within statistical precision}. The figure shows how the time-space coefficients oscillate in sign and take very large values in magnitude. In the eigen-space, they still sometimes change sign, and end up becoming as large as the ones in the time domain but, crucially, their absolute value changes almost monotonically and at a much slower rate. Either way, the large values of the coefficients result in uncontrollable statistical errors on $\rho_{\scriptstyle N}[\mathcal{S}]$. These, given the linearity of the problem, are given by the square root of the functional
\begin{flalign}
   B\left[\vec{g}_{\scriptscriptstyle [\mathcal{S}],N}\right] = 
   \vec{g}^T_{\scriptscriptstyle [\mathcal{S}],N} \cdot \matrix{Cov}_{\scriptstyle N} \cdot \vec{g}_{\scriptscriptstyle [\mathcal{S}],N} \,,
   \label{eq:B_functional}
\end{flalign}
where the positive-definite symmetric $N\times N$ matrix $\matrix{Cov}_{\scriptstyle N}$, with elements
\begin{flalign}
\left\langle \Delta_\mathrm{stat} C_{\scriptstyle N}(\tau n)\, \Delta_\mathrm{stat} C_{\scriptstyle N}(\tau m) \right\rangle\,,
\end{flalign}
is the statistical covariance of the correlator.

\begin{figure}[tb!]
    \centering
    \includegraphics[width=\globalWidth]{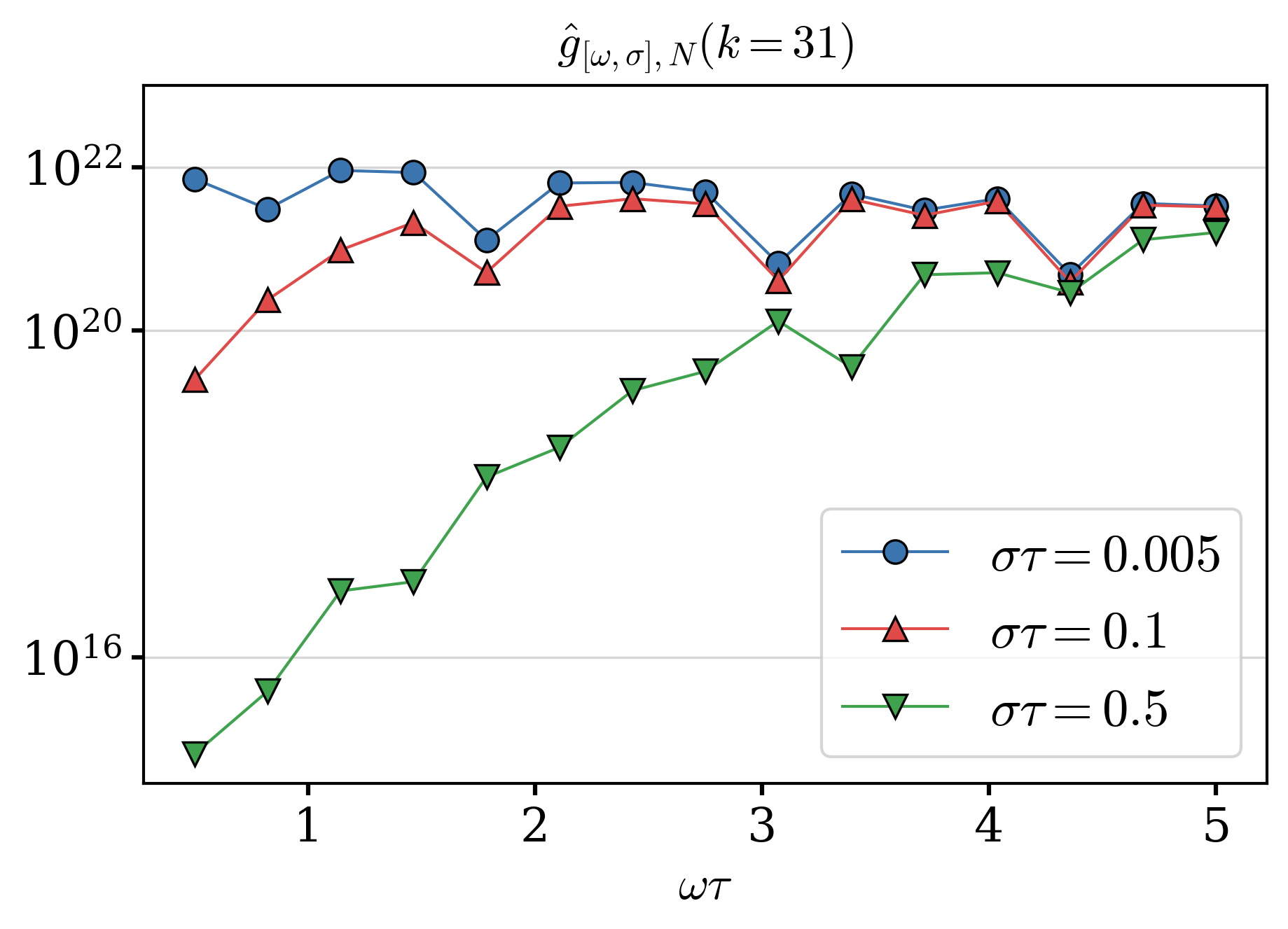}
    \caption{Eigen-space coefficient corresponding to a single eigenvalue ($k=31$, one of the last points in Fig.~\ref{fig:g_shared}) for Gaussian smearing kernels of different central values ($\omega$) and widths ($\sigma$). The size of the coefficients is descriptive of how ill-conditioned the problem is. This plot illustrates how this is related to the choice of smearing kernel. $\sigma \tau = 0.5$ (green) is a typically encountered value, while $\sigma \tau = 0.005$ (blue) is very extreme. The size is $N=32$.}
    \label{fig:g_change_omega_sigma}
\end{figure}

Fig.~\ref{fig:g_change_omega_sigma} illustrates the dependence of the coefficients on the smearing kernel. As it can be seen, reducing the width of the Gaussian, $\sigma$, or probing higher energies, $\omega$, results in larger coefficients, thus exacerbating the ill-conditioning of the problem. Conversely, the difficulty of the problem can be partially mitigated if a given smearing kernel is such that the associated vector $\vec{\hat{f}}_{\scriptstyle [\mathcal{S}],N}$ has a small overlap with the eigenvectors associated to small eigenvalues of $\matrix{A}_{\scriptstyle N}$.

\begin{figure}[tb!]
    \centering
    \includegraphics[width=\globalWidth]{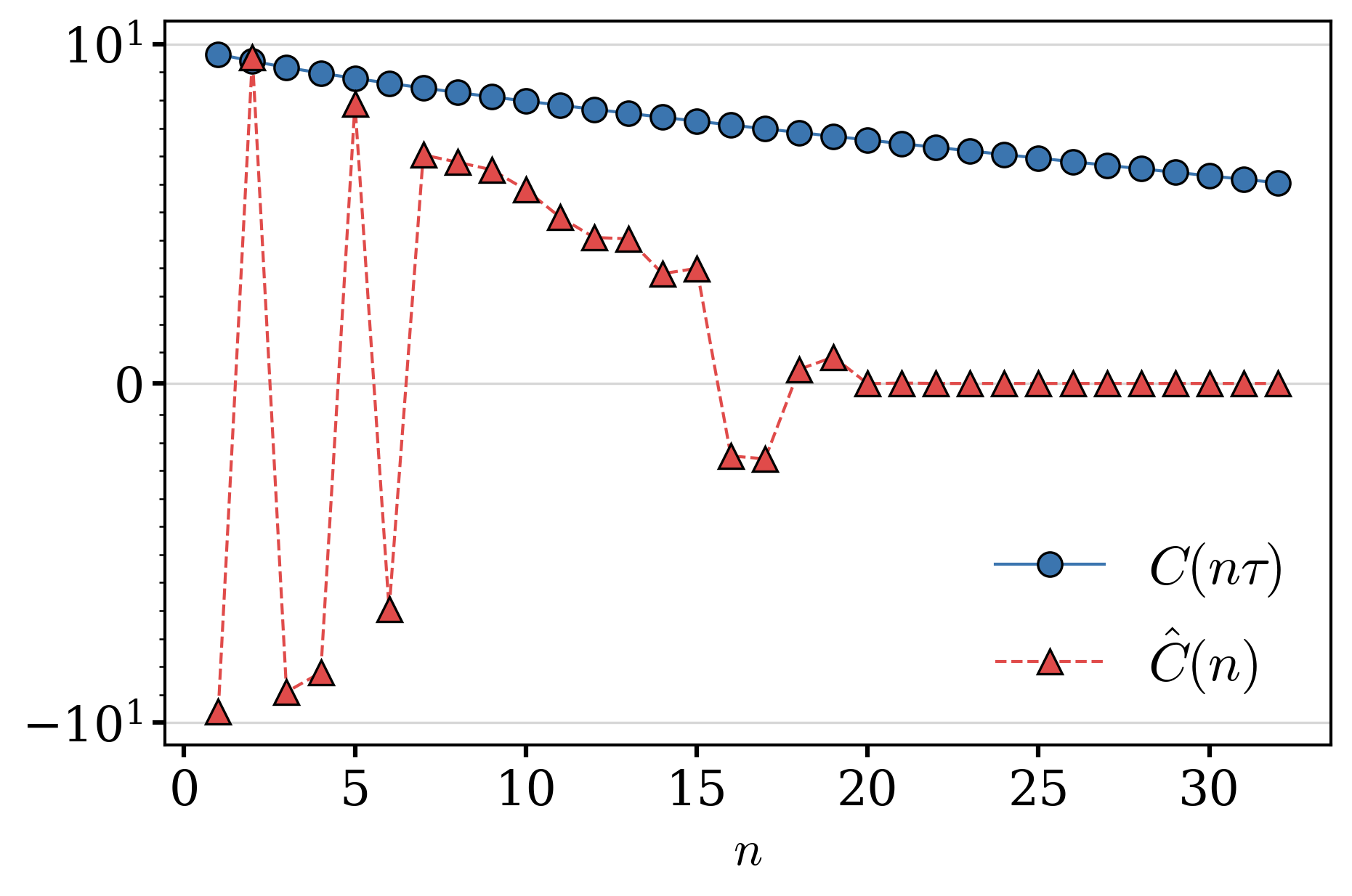}
    \caption{An instance of synthetic correlator used in this work (see Eqs.~\eqref{eq:mockrho} and~\eqref{eq:mockC}), in both the time and eigen-basis. This example, used to reconstruct the smeared spectral density in Fig.~\ref{fig:exact_sums_partial}, shows that the correlator in the eigen-basis decays quicker, partially mitigating the ill-conditioning of the problem when the representation of  $\hat g_{\scriptscriptstyle [\mathcal{S}],N}(n) \hat C(n)$ given in Eq.~\eqref{eq:rho_dual} is used. Details about the generation of synthetic correlators can be found in the main text and in Ref.~\cite{DelDebbio:2024lwm}.}
    \label{fig:C_and_Chat}
\end{figure}
\begin{figure}[t!]
    \centering
    \includegraphics[width=\globalWidth]{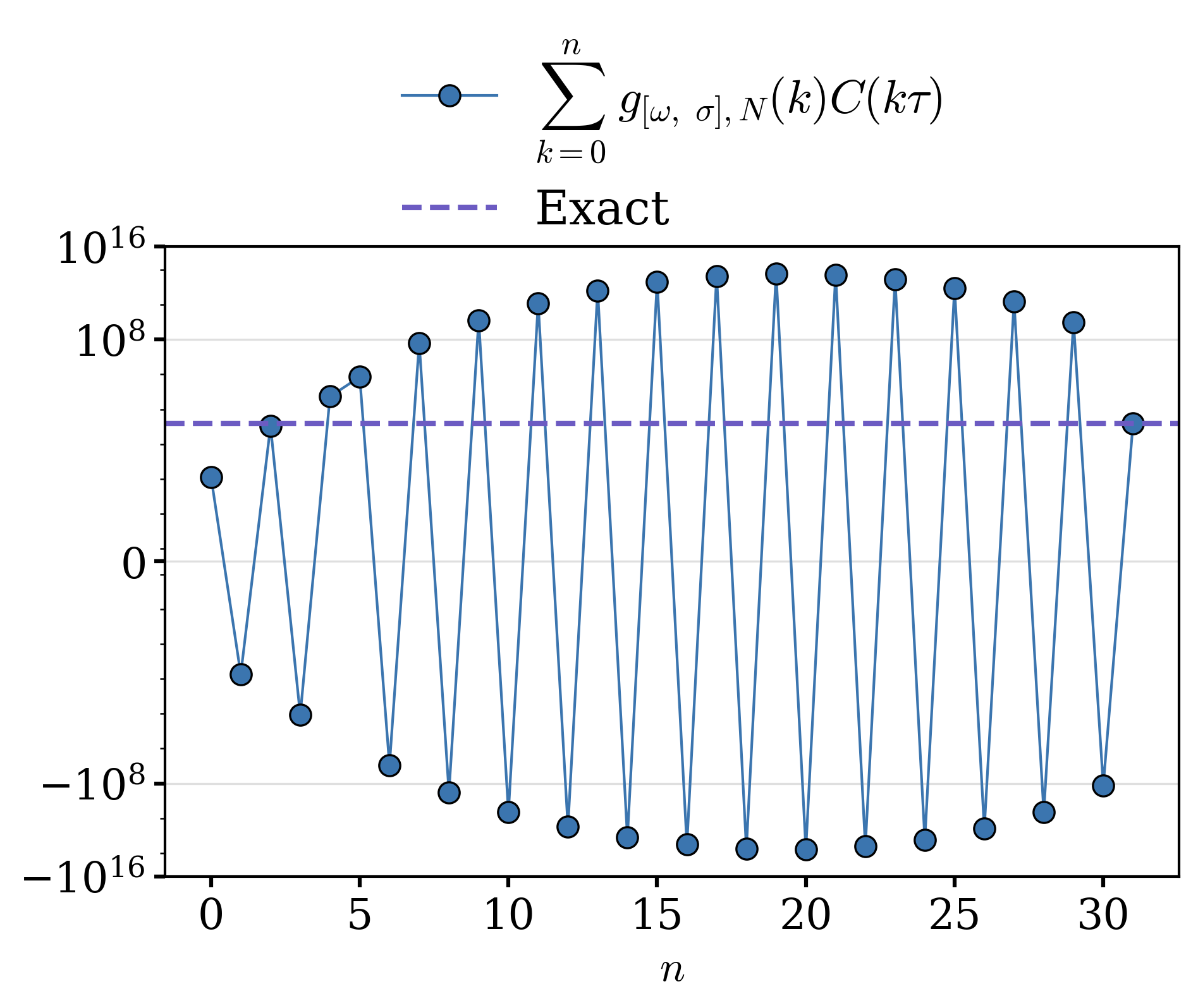}
    \includegraphics[width=\globalWidth]{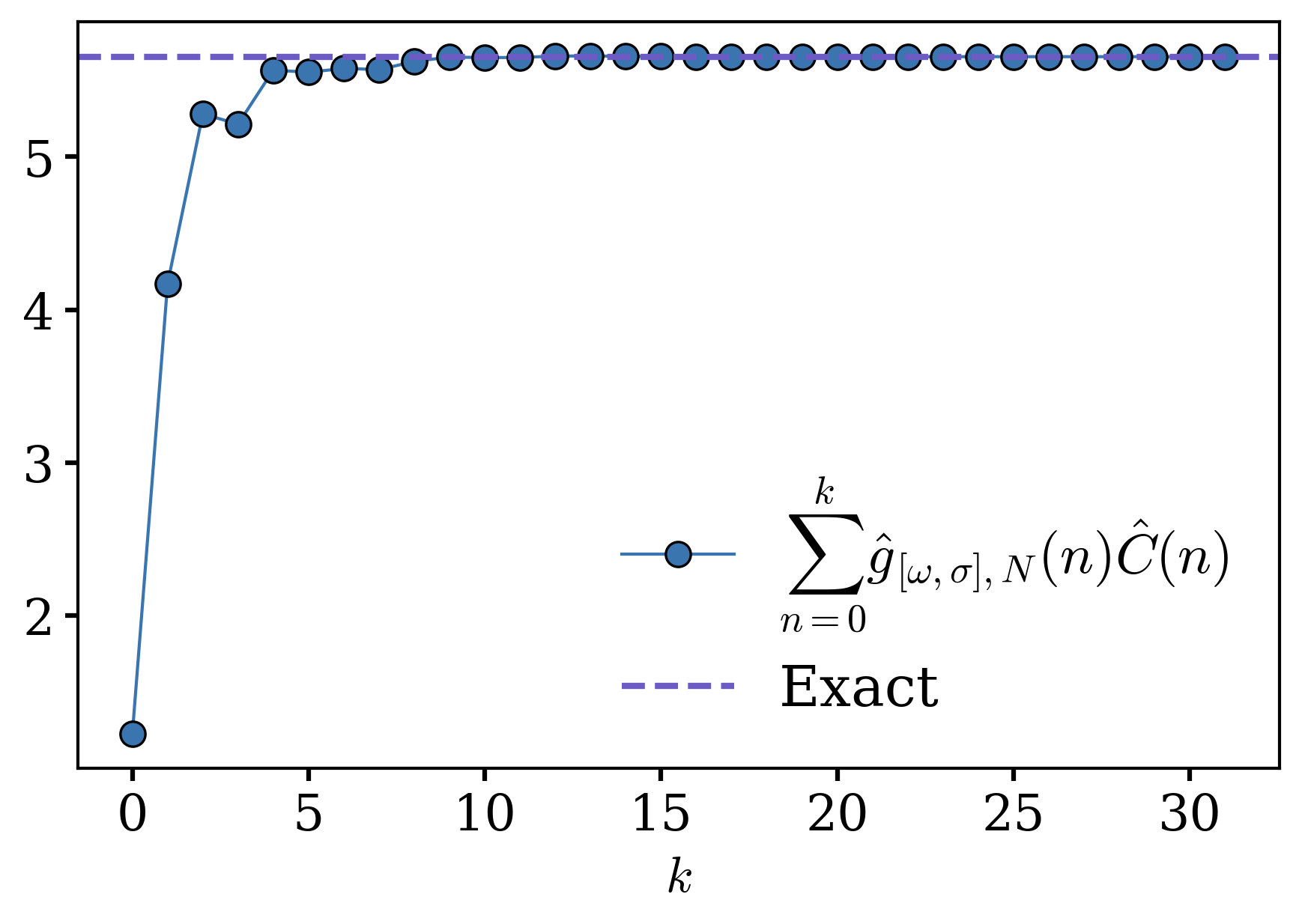}
    \caption{Partial sums leading to the smeared spectral density by summing terms in the time-domain (top) or eigen-domain (bottom), see Eqs.~\eqref{eq:rho_equals_C_dot_A_dot_f} and~\eqref{eq:rho_dual}, using the correlator from Fig.~\ref{fig:C_and_Chat}. The smearing kernel is a Gaussian with $\omega\tau=0.75$, $\sigma \tau =0.5$. The correlator has $N=32$ points. Summing over Euclidean times involves massive cancellations (top) and all terms in the sum must be included to get the correct result. Different is the case when the sum is written in terms of eigenvectors of $\matrix{A}_{\scriptstyle N}$ (bottom): the sum saturates after a few contributions. This feature will be central in the regularisation strategy proposed in Section~\ref{sec:decoupling_signal_from_noise}.}
    \label{fig:exact_sums_partial}
\end{figure}
To conclude this section, we show how the partial sums which lead to the smeared spectral density behave in the two domains. To do so, we introduce a first synthetic correlator, which in this section we maintain free from statistical noise. All the synthetic correlators used in this work are generated to represent a realistic discrete, finite-volume QCD spectrum. As already done in Ref.~\cite{DelDebbio:2024lwm}, we consider mock spectral densities of the form
\begin{flalign}
\rho(E) = \sum_{m=0}^{m_\mathrm{max}-1} w_m\, \delta(E-E_m)\;,
\quad
E_0 < E_1 \le \cdots \,,
\label{eq:mockrho}
\end{flalign}
and the corresponding correlators
\begin{flalign}
C(n\tau) = \sum_{m=0}^{m_\mathrm{max}-1} w_m\, e^{-n\tau E_m}\;,
\label{eq:mockC}
\end{flalign}
with $m_\mathrm{max}=O(10)$, and extract random values for the energies $\tau E_m$ in the range $[2\tau m_\pi,8\tau m_\pi]$, where $m_\pi$ is the physical value of the neutral pion mass. The spectral weights $w_m$ are extracted from a multivariate Gaussian distribution as described in detail in Ref.~\cite{DelDebbio:2024lwm} (see in particular Section 6, Eqs.~(51)--(53)).

An instance of synthetic correlator is shown in Fig.~\ref{fig:C_and_Chat} in both the time-space (blue) and the eigen-space (red). The corresponding partial sums, obtained by truncating at $k\le N$ respectively the time-space and the eigen-space representations of $\rho_{\scriptstyle N}[\mathcal{S}]$ given in Eqs.~\eqref{eq:rho_equals_C_dot_A_dot_f} and~\eqref{eq:rho_dual}, are shown in Fig.~\ref{fig:exact_sums_partial}. In the time-space representation all contributions to the smeared spectral density are important and the final result comes out from fine-tuned cancellations once \emph{all} the terms of the sum are included. In the eigen-space, on the other hand, the terms $\hat g_{\scriptscriptstyle [\mathcal{S}],N}(n) \hat C(n)$ corresponding to the smaller eigenvalues of $\matrix{A}_N$ become negligible and the sum saturates for $k\ll N$. This will carry important consequences when we will add statistical noise to the input data.

In the next section we discuss two regularizations of the problem, together with the associated {\color{MODCOLOR} numerical extrapolation} procedures to remove the regulators and to estimate the systematic and statistical errors on $\rho[\mathcal{S}]$: the well assessed ``stability analysis'', and a new method, that we call the ``eigen-space analysis'', which allows to partially decouple the signal from the statistical noise.

%% file: sections/newmethod.tex
\section{Regularisations}\label{sec:regularisations}

\subsection{Stability Analysis}\label{sec:stability}

In the HLT method the problem is naturally approached in time-space, the regulator is chosen to be the one prescribed by Backus and Gilbert~\cite{10.1111/j.1365-246X.1968.tb00216.x}, {\color{MODCOLOR}and the numerical procedure to estimate the associated systematic error} goes under the name of stability analysis (SA).

The coefficients $\vec{g}_{\scriptscriptstyle [\mathcal{S}],N}$ are regulated by modifying Eq.~\eqref{eq:g_as_argmin_of_A} to be
\begin{flalign}
    \vec{g}_{\scriptscriptstyle [\mathcal{S}],N,\lambda} = \underset{\vec g \in \mathbb{R}^N}{\mathrm{argmin}} \; \left\{ A[\vec g] + \lambda B[\vec {g}] \,  \right\} \, ,
    \label{eq:g_as_argmin_of_A_p_B}
\end{flalign}
where $B[\vec g]\ge 0$ is the error functional introduced in Eq.~\eqref{eq:B_functional} and $\lambda\ge 0$ is the so-called trade-off parameter. The regulated versions of Eq.~\eqref{eq:g_equals_A_dot_f} and Eq.~\eqref{eq:rho_equals_C_dot_A_dot_f} are then
\begin{flalign}
&
    \vec{g}_{\scriptscriptstyle [\mathcal{S}],N,\lambda} = 
    \frac{1}{\matrix{A}_N +\lambda \matrix{Cov}_N}\cdot 
    \vec{f}_{\scriptscriptstyle [\mathcal{S}],N}\,,
\nonumber \\[8pt]
&
    \rho_{\scriptstyle N,\lambda}[\mathcal{S}] = \vec{C}_N^T \cdot 
    \frac{1}{\matrix{A}_N +\lambda \matrix{Cov}_N}\cdot 
    \vec{f}_{\scriptscriptstyle [\mathcal{S}],N}\,.
    \label{eq:glambda}
\end{flalign}
The virtue of the Backus-Gilbert regulator is that of removing the quasi-singularity associated with the small eigenvalues of the matrix $\matrix{A}_N$ by adding a term that is proportional to the statistical error of the correlator. This important fact makes the Backus-Gilbert regulator \textcolor{MODCOLOR2}{asymptotically} unbiased, {\color{MODCOLOR} in the sense that} in the ideal limit in which the correlator is exact, the covariance matrix and the $B$ functional vanish identically and the very same result {\color{MODCOLOR} $\rho_{\scriptstyle N}[\mathcal{S}]$} is obtained for any value of $\lambda$, including zero. The price to pay is an error-dependent distortion of the smearing kernel and the consequent dependence of $\rho_{\scriptstyle N,\lambda}[\mathcal{S}]$ upon $\lambda$. The physical result $\rho[\mathcal{S}]$ must therefore be obtained by taking the limits\footnote{{\color{MODCOLOR} The Tikhonov regularisation, thoroughly discussed in~\cite{Bertero_1988}, amounts to substitute the $B[\vec g]$ functional with the error-independent functional $\lambda \vec g^T\cdot \vec g$. The dependence upon $\lambda$ of the resulting smearing kernel is then predictable, and in Ref.~\cite{Bruno:2024fqc} has been profitably used as a handle, but it doesn't automatically disappear in the limit of vanishing statistical errors, as instead happens in the unbiased Backus-Gilbert case.}}
\begin{flalign}
\rho[\mathcal{S}]=\lim_{N\mapsto \infty}\lim_{\lambda\mapsto 0} \rho_{\scriptstyle N,\lambda}[\mathcal{S}]\, . 
\label{eq:Nlambda_limits}
\end{flalign}

{\color{MODCOLOR}
The HLT stability analysis~\cite{Hansen:2019idp,Bulava:2021fre} is a numerical procedure to take these limits {\color{MODCOLOR2} by checking for convergence within statistical precision}. It is built upon the numerical fact that lattice results are unavoidably affected by statistical errors and upon the theoretical fact that the limits of Eq.~\eqref{eq:Nlambda_limits} exist, so that the systematic error $\left\vert \rho[\mathcal{S}]-\rho_{\scriptstyle N,\lambda}[\mathcal{S}]\right\vert$ can be made arbitrarily small by increasing $N$ and by decreasing $\lambda$.  The goal is that of obtaining a numerical estimate $\rho_\star[\mathcal{S}]$ of $\rho[\mathcal{S}]$ such that the residual systematic error (which is unavoidable, being this a numerical extrapolation)  is smaller than the statistical error. It is worth stressing that the threshold value $N_\star$ such that the results stabilize within statistical errors for $N>N_\star$ depends on the physical application and on the target total error. This means that if the target accuracy is too high one might need values of $N$ which are prohibitively large from the lattice perspective. Moreover, in the SA the residual systematic error is estimated numerically and, from a rigorous mathematical perspective, observing stability of the results within the simulated range of $N$ doesn't necessarily imply that significant variations cannot show up at larger values of $N$.

The systematic error can be split in two contributions,
\begin{flalign}
\left\vert 
\rho[\mathcal{S}]
-
\rho_{\scriptstyle N,\lambda}[\mathcal{S}]
\right\vert
\le
\left\vert 
\rho[\mathcal{S}]
-
\rho_{\scriptstyle N}[\mathcal{S}]
\right\vert
+
\left\vert 
\rho_{\scriptstyle N}[\mathcal{S}]
-
\rho_{\scriptstyle N,\lambda}[\mathcal{S}]
\right\vert \;.
\label{eq:sys_split}
\end{flalign}
The fact that the first contribution on the r.h.s. can be made arbitrarily small by increasing $N$ is a direct consequence of the Weierstrass theorem. This can easily be understood in the cases in which $\rho(E)$ is a $\mathcal{L}_2$-tempered function, e.g.\ by considering $\alpha>0$ and the following bound
\begin{flalign}
&
\left\vert 
\rho[\mathcal{S}]
-
\rho_{\scriptstyle N}[\mathcal{S}]
\right\vert
\nonumber \\[8pt]
&
=
\left\vert 
\int_{E_0}^{\infty} dE \, 
e^{\frac{\alpha\tau E}{2}}
\left[\mathcal{S}(E) - \mathcal{S}_{\scriptstyle N}(E)\right]\, 
e^{\frac{-\alpha\tau E}{2}}\, \rho(E)
\right\vert
\nonumber \\[8pt]
&
\le
\sqrt{A[\vec{g}_{\scriptscriptstyle [\mathcal{S}],N}]}
\,
\left\|
\rho(E)
\right\|_{-\alpha} \;.
\label{eq:sys_firtsterm_L2}
\end{flalign}
Indeed, provided that the integral 
\begin{flalign}
&
\left\|
\rho(E)
\right\|^2_{-\alpha}
=
\int_{E_0}^{\infty} dE \, 
e^{-\alpha\tau E}\, \rho^2(E)
\end{flalign}
exists and is finite, Eq.~\eqref{eq:sys_firtsterm_L2} shows that the contribution $\left\vert \rho[\mathcal{S}]-\rho_{\scriptstyle N}[\mathcal{S}]\right\vert$ to the systematic error is $O(\sqrt{A[\vec{g}_{\scriptscriptstyle [\mathcal{S}],N}]})$. In the cases in which $\left\|\rho(E)\right\|_{-\alpha}$ cannot be defined one can e.g.\ consider the following bound
\begin{flalign}
\left\vert 
\rho[\mathcal{S}]
-
\rho_{\scriptstyle N}[\mathcal{S}]
\right\vert
\le
C_+(\alpha \tau)\,
\left\|
e^{\alpha\tau E}
\left[\mathcal{S}(E) - \mathcal{S}_{\scriptstyle N}(E)\right]\,
\right\|_{\infty}
 \;,
\label{eq:sys_firtsterm_Linf}
\end{flalign}
where 
\begin{flalign}
C_+(\alpha \tau)
=
\int_{E_0}^{\infty} dE \, 
e^{-\alpha\tau E}\, \vert \rho(E) \vert\;,
\end{flalign}
which for positive spectral densities is the Euclidean correlator $C(\alpha \tau)$  at time $\alpha\tau$, and
\begin{flalign}
\left\|
f(E)
\right\|_{\infty} = \max_{E\in [E_0,\infty]} \left\vert f(E)\right\vert\;.
\end{flalign}
In this case too, the fact that $\left\vert \rho[\mathcal{S}]-\rho_{\scriptstyle N}[\mathcal{S}]\right\vert$ can be made arbitrarily small by increasing $N$ follows from the Weierstrass theorem. \textcolor{MODCOLOR}{A detailed analysis in the case of generic multivariate spectral densities has been performed in Ref.~\cite{Patella:2024cto}. There, by exploiting the fact that any tempered distribution can be written as the sum of possibly higher-order distributional derivatives of $\mathcal{L}_2$-tempered functions, the use of the $\left\| \cdot \right\|_{\infty}$ norm has been avoided}.

We now consider the second contribution to the systematic error in Eq.~\eqref{eq:sys_split}.} It is a simple exercise to show that Eq.~\eqref{eq:glambda}  implies
\begin{flalign}
&
\rho_{\scriptstyle N,\lambda_{2}}[\mathcal{S}]
-
\rho_{\scriptstyle N,\lambda_{1}}[\mathcal{S}]
\nonumber \\[8pt]
&
=
(\lambda_{1}-\lambda_{2})
\vec{C}_N^T \cdot
\frac{1}{\matrix{A}_N +\lambda_1 \matrix{Cov}_N}
\cdot
\matrix{Cov}_N
\cdot
\vec{g}_{\scriptscriptstyle [\mathcal{S}],N,\lambda_{2}}\,,
\end{flalign}
 from which it follows that
\begin{flalign}
&
\left\vert
\rho_{\scriptstyle N,\lambda}[\mathcal{S}]
-
\rho_{\scriptstyle N}[\mathcal{S}]
\right\vert
\nonumber \\[8pt]
&
\le
\lambda\, 
\left\| \sqrt{\matrix{Cov}_N}\cdot  \vec{g}_{\scriptscriptstyle [\mathcal{S}],N,\lambda}\right\|\,
\left\| \sqrt{\matrix{Cov}_N}\cdot \matrix{A}_N^{-1}\cdot \vec{C}_N \right\|
\,,
\end{flalign}
and therefore that, {\color{MODCOLOR} asymptotically and for each finite value of $N$}, the corrections to the $\lambda\mapsto 0$ limit  are $O(\lambda \sqrt{B[\vec{g}_{\scriptscriptstyle [\mathcal{S}],N,\lambda}]})$.

\begin{figure}[tb]
    \centering
    \includegraphics[width=\globalWidth]{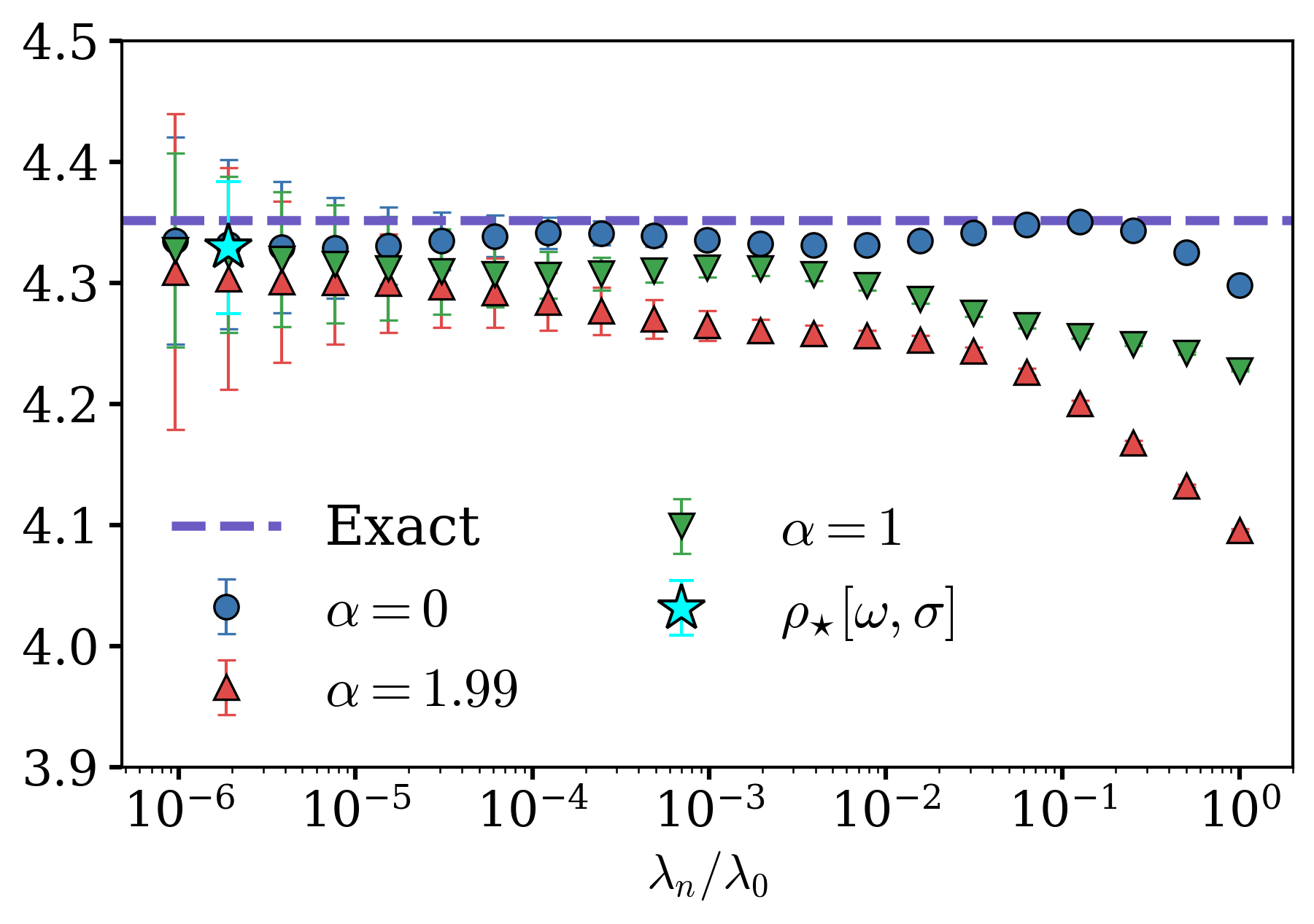}
    \includegraphics[width=\globalWidth]{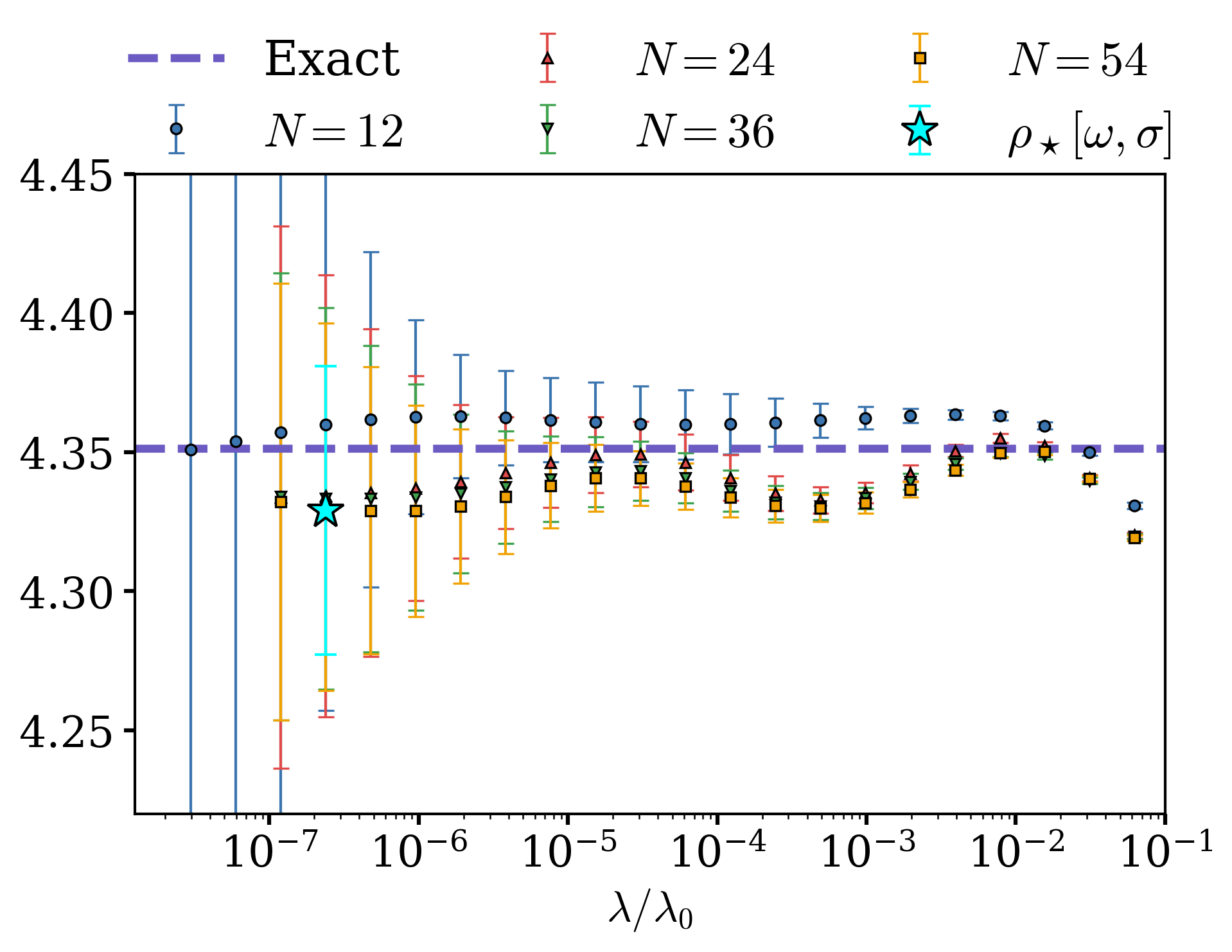}
    \caption{Top: example of stability analysis to remove the dependence upon the Backus-Gilbert regulator, parametrised here by $\lambda_n / \lambda_0$, performed using three values of $\alpha$ to reconstruct the smeared density at $\omega= m_\rho$, with $\sigma=2m_\pi$. Bottom: the same stability analysis, when performed for different values of $N$, allows to take the $N\rightarrow \infty$ limit, as described in the main text. In this last plot, we only display results at $\alpha=0$ for visual clarity. The \textcolor{turquoise}{turquoise} point $\rho_{\star}[\omega,\sigma]$ represents the result of the limits $\lim_{N \rightarrow \infty} \lim_{\lambda \rightarrow 0} \rho_{N,\lambda}[\omega,\sigma]$ within its quoted error bar.}\label{fig:SA_example}
\end{figure}

{\color{MODCOLOR} The SA is built on the previous theoretical observations}. It starts from a value of $\lambda$, let's call it $\lambda_\mathrm{0}$, such that the statistical error on $\rho_{\scriptstyle N,\lambda}[\mathcal{S}]$ is under control while the approximation of the smearing kernel is possibly rather bad, i.e.\ $A[\vec{g}_{\scriptscriptstyle [\mathcal{S}],N,\lambda_{0}}]\gg B[\vec{g}_{\scriptscriptstyle [\mathcal{S}],N,\lambda_{0}}]$. The value of $\lambda$ is then progressively reduced as shown in the top panel of Fig.~\ref{fig:SA_example}, e.g.\ by choosing $\lambda_n= 2^{-n}\, \lambda_0$. The procedure ends at $\lambda_\star\equiv \lambda_{n_\star}$ when the result $\rho_\star[\mathcal{S}]\equiv\rho_{\scriptstyle N,\lambda_\star}[\mathcal{S}]$ is such that
\begin{flalign}
&
A[\vec{g}_{\scriptscriptstyle [\mathcal{S}],N,\lambda_{n_\star}}]
\ll 
B[\vec{g}_{\scriptscriptstyle [\mathcal{S}],N,\lambda_{n_\star}}]\,,
\nonumber \\[8pt]
&
\left\vert
\rho_{\scriptstyle N,\lambda_{n_\star+1}}[\mathcal{S}]
-
\rho_\star[\mathcal{S}]
\right\vert^2
\ll 
B[\vec{g}_{\scriptscriptstyle [\mathcal{S}],N,\lambda_{n_\star}}]\,,
\end{flalign}
i.e.\ when the systematic error on $\rho_\star[\mathcal{S}]$ can safely be neglected w.r.t.\ its statistical error, which usually keeps increasing rather fast by decreasing $\lambda$. \textcolor{MODCOLOR2}{This allows to numerically take the limit $\lambda \rightarrow 0$ within statistical errors, at fixed $N$.}

The $N\mapsto \infty$ limit is performed analogously. The numerical onset of the limit is reached  when, for $\lambda$ sufficiently small and $N$ sufficiently large, the residual dependence of $\rho_{\scriptstyle N,\lambda}[\mathcal{S}]$ upon $N$ is negligible w.r.t.\ the statistical error. This is shown in the bottom panel of Fig.~\ref{fig:SA_example}. Moreover, by relying upon the fact that Eq.~\eqref{eq:exact_representation_of_kernel} holds for any value of $\alpha<2$ in Eq.~\eqref{eq:g_as_argmin_of_A}, and that therefore any dependence upon $\alpha$ of $\rho_{\scriptstyle N,\lambda}[\mathcal{S}]$ must be due to the fact that $\lambda>0$ and/or $N<\infty$, one can {\color{MODCOLOR} approach both limits from different numerical paths, corresponding to different values of $\alpha$. This gives the possibility to strengthen the numerical check that the onset of the $\lambda\mapsto 0$ and $N\mapsto \infty$ limits has been reached {\color{MODCOLOR2}  within statistical errors}. An example of the limit $\lambda \rightarrow 0$ being reached \textcolor{MODCOLOR2}{in this way} using different values of $\alpha$ as a cross-check, is shown in the top panel of Fig.~\ref{fig:SA_example}}.

{\color{MODCOLOR2}
We want to stress that in the SA the $\lambda\mapsto 0$ and $N\mapsto \infty$ limits are taken without attempting to fit their corrections. Modeling these corrections is possible, in principle, and it would allow to reduce the systematic errors associated with the numerical extrapolations. Actually, since statistical errors blow up by reducing $\lambda$ and increasing $N$, the choice of considering the simplest possible model, i.e.\ the constant function, and to take the numerical limits by searching for a plateau within statistical errors, is the one that provides the more conservative estimate of the total error.
}

{\color{MODCOLOR}
At the end of the SA, $\rho_\star[\mathcal{S}]$ is an estimate of $\rho[\mathcal{S}]$ which has been proven to be robust in several cases: in support of this statement, we report again studies within the integrable (and therefore analytically-controllable) two-dimensional $O(3)$ non-linear  $\sigma$-model, a number of closure tests for lattice QCD~\cite{Buzzicotti:2023qdv, DelDebbio:2024lwm}, more of which will be illustrated in Section~\ref{sec:pulls}, and tests against other spectroscopy methods~\cite{DelDebbio:2022qgu, Bennett:2024cqv}. While these tests provide quantitative evidence of the robustness of this procedure, it should be stressed that the SA is a numerical extrapolation. The study of the dependence of the results on $N$ and on the trade-off parameter $\lambda$ is a delicate task and must be done with the due care. Eqs.~\eqref{eq:sys_firtsterm_L2} and~\eqref{eq:sys_firtsterm_Linf} are rigorous a priori bounds but the SA analysis doesn't exclude the possibility that significant corrections might show up beyond the numerically accessible range of $N$.
{\color{MODCOLOR2}
To the light of this observation, we present in the next subsection an alternative way of regularizing and controlling the finite-$N$ reconstruction. \textcolor{MODCOLOR3}{Through the introduction of an eigenspace cutoff on the matrix $\matrix{A}_N$ in Eq.~\eqref{eq:Au_au}, this new procedure manifestly decouples part of the noise from the signal in Eq.~\eqref{eq:rho_as_sum_g_dot_c} without the introduction of the Backus-Gilbert regulator.}}
}

\subsection{Eigen-space Analysis}\label{sec:decoupling_signal_from_noise}

In this subsection we introduce the new procedure, the main novelty of this work, that we call eigen-space analysis (EA). We want to stress that what we consider a novelty is the algorithmic procedure to estimate {\color{MODCOLOR} numerically} the statistical and systematic errors on smeared spectral densities. Indeed, as we already remarked in the introduction, the mathematics to approach this class of problems is well known and well established.

The idea is essentially the same behind the Singular Value Decomposition approach to the inversion of (nearly) singular matrices: the solution is found in the subspace orthogonal to that of the nearly-vanishing eigenvalues of the matrix. This approach is particularly useful {\color{MODCOLOR} to get a numerical estimate} of $\rho[\mathcal{S}]$ because it allows one to partially decouple the signal from the noise: it turns out that the contributions to $\rho[\mathcal{S}]$ coming from the larger (smaller) eigenvalues of $\matrix{A}_N$ mostly contribute to the signal (noise). As a consequence, the terms which carry most of the noise give irrelevant contributions to the central value.

While the SA analysis procedure is naturally formulated in the time-space, where $\rho_N[\mathcal{S}]$ is given by the sum of Eq.~\eqref{eq:rho_as_sum_g_dot_c}, the starting point of the new procedure is the eigen-space representation given in Eq.~\eqref{eq:rho_dual}. {\color{MODCOLOR2} Without statistical errors, the behaviour of the partial sums
\begin{flalign}
 \rho_{\scriptstyle N}^{\scriptstyle k}[\mathcal{S}]
 =\sum_{n=1}^k \hat{g}_{\scriptstyle [\mathcal{S}],N}(n) \hat{C}_{\scriptstyle N}(n)\,,
\end{flalign}
such that 
\begin{flalign}
&
 \rho_{\scriptstyle N}[\mathcal{S}] \equiv  \rho_{\scriptstyle N}^{\scriptstyle N}[\mathcal{S}]\;,
 \nonumber \\[8pt]
 &
\textcolor{MODCOLOR3}{\rho[\mathcal{S}]  = \lim_{N\mapsto\infty} \lim_{k\mapsto N} \rho_{\scriptstyle N}^{\scriptstyle k}[\mathcal{S}]}
= \lim_{N\mapsto\infty}  \rho_{\scriptstyle N}^{\scriptstyle N}[\mathcal{S}]\,,
    \label{eq:rho_dual2}
\end{flalign}
was illustrated in the bottom panel of Fig.~\ref{fig:exact_sums_partial}, and it is radically different from the time-domain partial sums shown in the top panel}. The results become even more interesting in the presence of noise. To illustrate this point we use one of our mock correlators introduced in Section~\ref{sec:theo_frame}, and we inject noise taken from a covariance matrix measured on the lattice for a vector-vector correlator (see also Fig.\ref{fig:noisy_corr_example}). 

\begin{figure}
    \centering
    \includegraphics[width=\globalWidth]{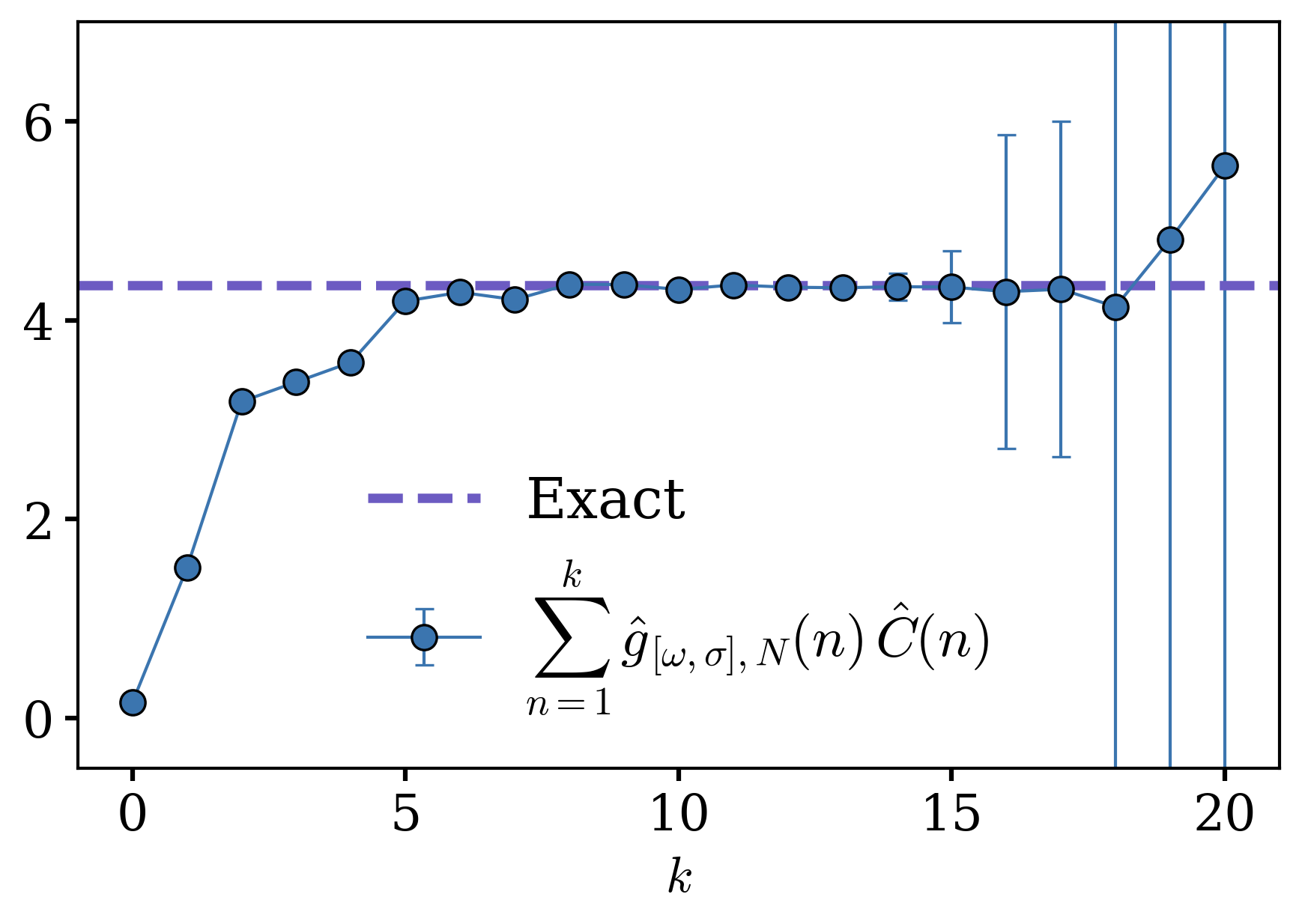}
    \includegraphics[width=\globalWidth]{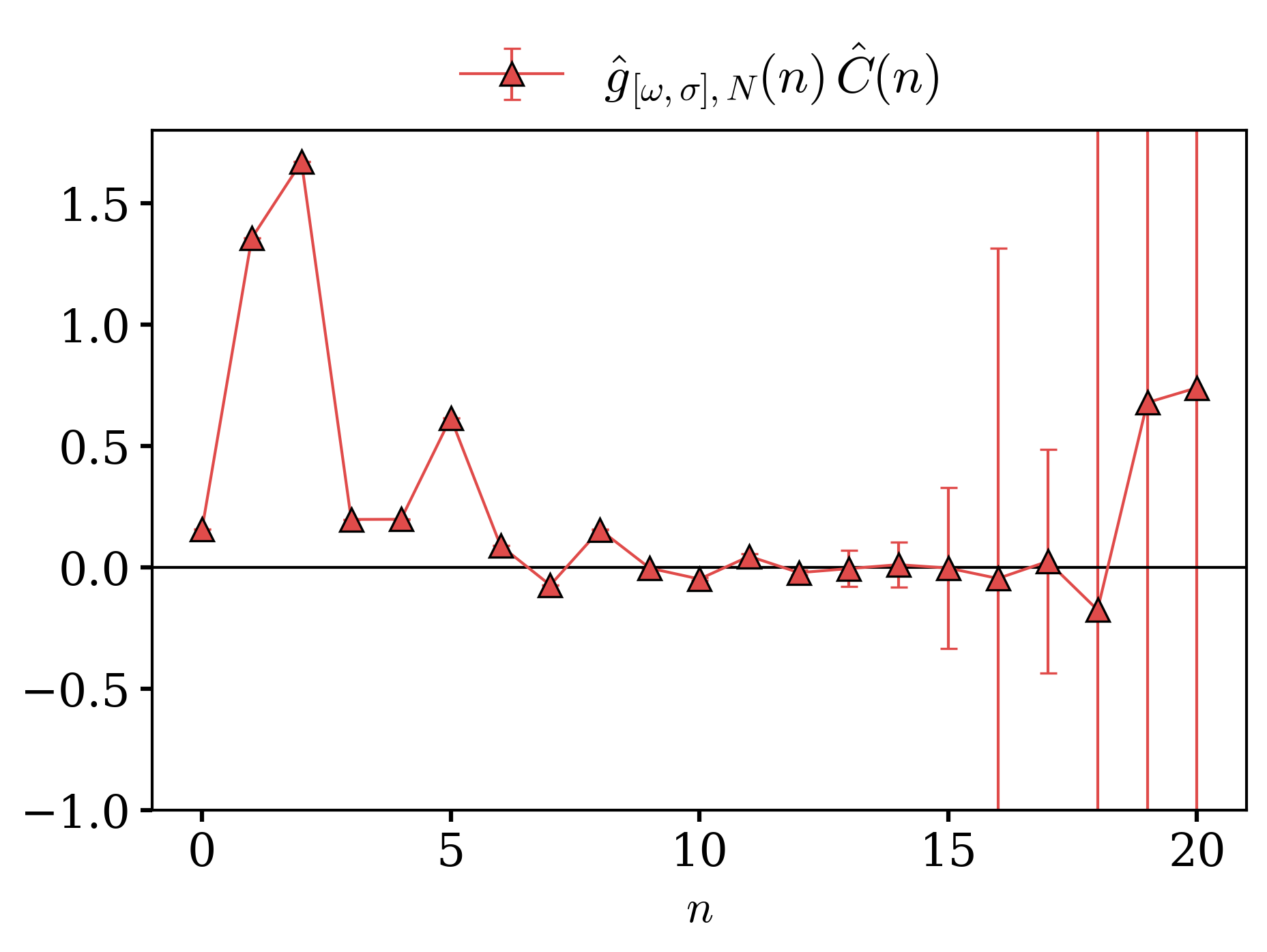}
    \caption{Smeared spectral density from the sum in Eq.~\eqref{eq:rho_dual}, reconstructed at $\omega=m_\rho$ with $\sigma = 2 m_\pi$, with a mock vector-vector correlator affected by a realistic statistical noise. Top: partial sum. Bottom: individual terms. The key observation is that the signal saturates before the noise reaches its maximum level, suggesting the possibility to truncate the sum.}
    \label{fig:k_terms_noisy}
\end{figure}

The results are displayed in Fig.~\ref{fig:k_terms_noisy}, showing individual terms (bottom) and the partial sum (top) of Eq.~\eqref{eq:rho_dual}. The terms contributing the most to the signal, corresponding to the larger eigenvalues $a_N(k)$, have lower statistical error. Conversely, the terms associated to smaller eigenvalues contribute progressively less to the signal, while they grow noisier. There seems to be a window where the sum saturates before the noise explodes, suggesting the possibility to truncate the sum. Compared to the SA, where one looks for a plateau in the bias parameter $\lambda$ before the noise explodes, here one has to find a plateau in the signal and truncate it, again, before the noise takes over.

In the next Section, we will define an algorithmic procedure for truncating the sum over eigenstates, test its performance and compare it to that of the SA.

%% file: sections/pulls.tex
\section{Algorithmic procedure and performance analysis}\label{sec:pulls}

\begin{figure}[tb]
    \centering
    \includegraphics[width=\globalWidth]{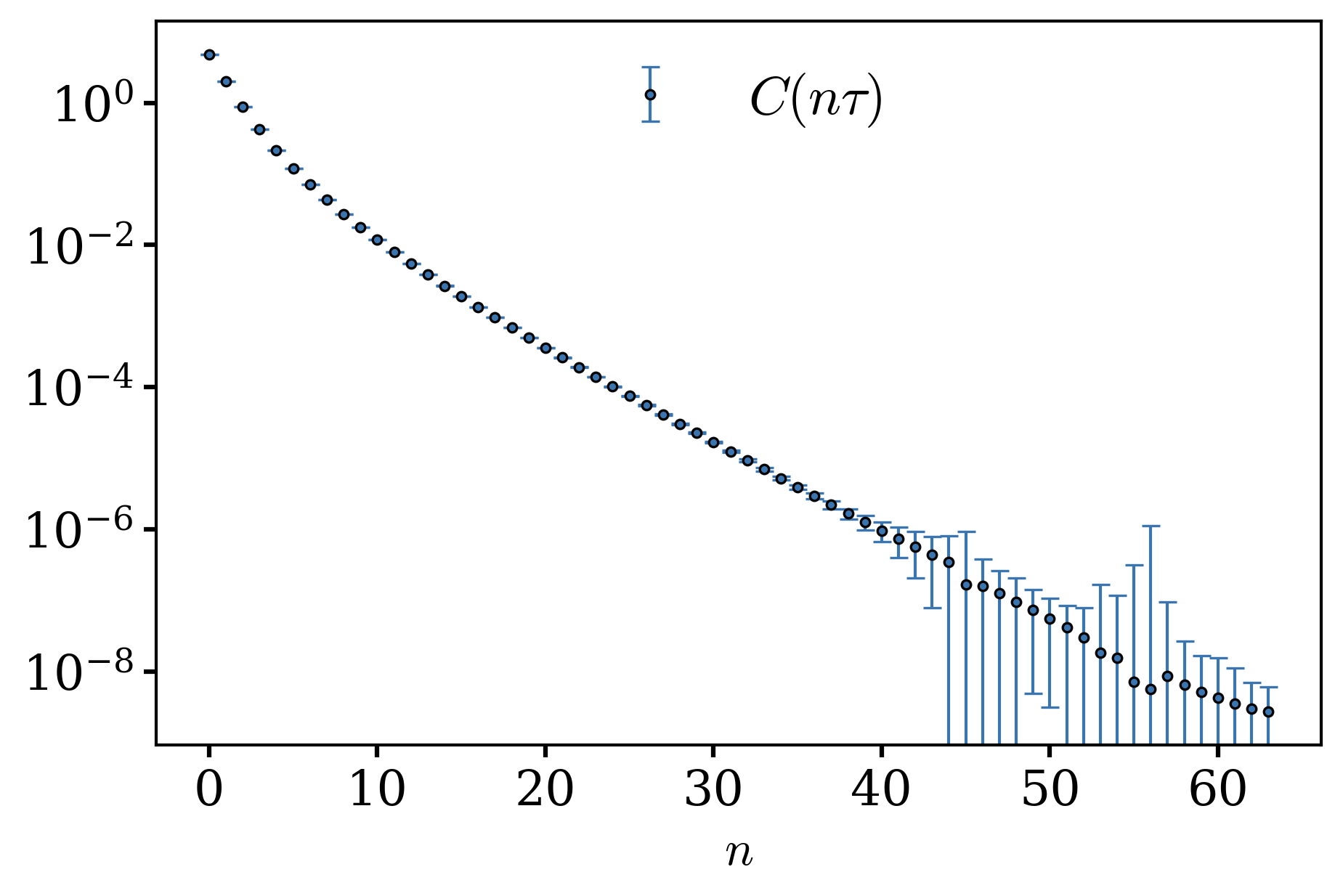}
    \includegraphics[width=\globalWidth]{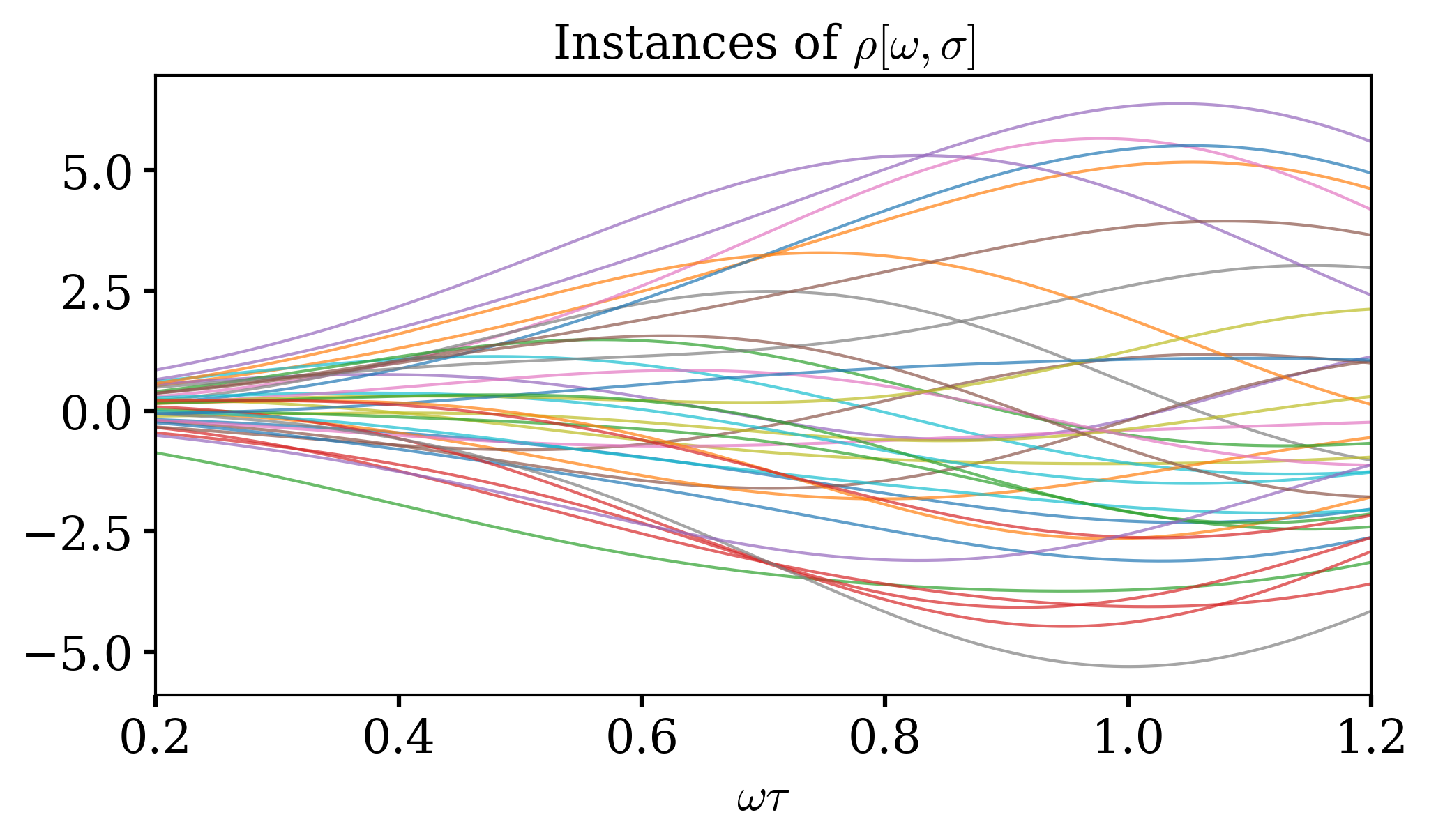}
    \caption{Top: single example of mock vector-vector correlator used to perform the closure tests. We generate over a thousand of such correlators, and we solve the inverse problem on each of them. Of these thousands, in the bottom panel, we display some examples of the exact smeared spectral densities we try to reconstruct from the various mock correlators when performing the closure tests. Here, $\sigma \tau = 0.27$.}\label{fig:noisy_corr_example}
\end{figure}

In this section we present an operative criterion for truncating the sum of Eq.~\eqref{eq:rho_dual} representing $\rho_N[\mathcal{S}]$ in eigen-space. In order to systematically assess the performance of the procedure, we perform closure tests, i.e.\ we solve the inverse problem for over a thousand different datasets. To this end, we generate sets of pairs of synthetic spectral densities and correlators according to Eqs.~\eqref{eq:mockrho} and~\eqref{eq:mockC} (see also Ref.~\cite{DelDebbio:2024lwm}). These are then scattered with statistical noise distributed according to a realistic lattice covariance for a vector-vector correlator \textcolor{MODCOLOR}{with $N=64$}, from a state-of-the-art lattice simulation\footnote{We thank F. Margari and the ETM collaboration for providing such covariance matrix.}. In the top panel of Fig.~\ref{fig:noisy_corr_example} we show one of the thousand correlators that we consider: \textcolor{MODCOLOR}{these have the same level of noise of those from which one can extract the $R$-ratio~\cite{ExtendedTwistedMassCollaborationETMC:2022sta} or the anomalous magnetic moment of the muon~\cite{Aliberti:2025beg}. In the bottom panel of Fig.~\ref{fig:noisy_corr_example} we show examples of the smeared spectral densities associated to some of the mock correlators. Notice that, in order to cover a broader class of physical applications w.r.t.\ the $R$-ratio, from which the statistical covariance comes from, these are allowed to change sign.}

{\color{MODCOLOR2}
We truncate the sum of Eq.~\eqref{eq:rho_dual} at $N_{\star\star}$, i.e.
\begin{flalign}
\rho_{\star\star}[\mathcal{S}] \equiv
\rho_{\scriptstyle N}^{\scriptstyle N_{\star\star}}[\mathcal{S}]
=
\sum_{n=1}^{N_{\star\star}} \hat g_{\scriptstyle [\mathcal S],N}(n) \hat C_{\scriptstyle N}(n)
\,,
\quad N_{\star\star} \le N\,,
\end{flalign}
when $n_{\rm stop}$ contributions, including the $N_{\star\star}$ one, are compatible with zero within their own statistical error, 
\begin{flalign}
&
\left\vert \hat g_{\scriptstyle [\mathcal S],N}(n) \hat C_{\scriptstyle N}(n) \right\vert
\le 
\Delta_\mathrm{stat} \left[
\hat g_{\scriptstyle [\mathcal S],N}(n) \hat C_{\scriptstyle N}(n)
\right]
\,,
\nonumber \\[8pt]
&
N_{\star\star}-n_{\rm stop} <n\le N_{\star\star}\,.
\end{flalign}
Also in this case, we don't attempt to model and fit the corrections to the $N_{\star\star}\mapsto \infty$ limit but we progressively increase $N$ and $N_{\star\star}$ and check for convergence within statistical errors}

We test the choices $n_{\rm stop}=1,2,3$ by solving the inverse problem over a thousand times and comparing each choice of truncation with the exact result. We consider the Gaussian kernel of Eq.~\eqref{eq:gaussian} with $\sigma = 2m_\pi = 0.27~\mbox{GeV}$ and we perform the reconstruction at the energy $\omega = m_\rho = 0.77~\mbox{GeV}$. Given also the nature of the synthetic correlators discussed above, these values are illustrative of a state-of-the-art reconstruction of the smeared R-ratio~\cite{ExtendedTwistedMassCollaborationETMC:2022sta}, and therefore allow us to test our procedure in a relevant situation. The outcome is given in terms of the pull variable
\begin{flalign}\label{eq:def_pull}
    p[\omega,\sigma] = \frac{\rho_{\star\star}[\omega,\sigma] - \rho_{\rm true}[\omega,\sigma]}{\Delta_{\rm stat}\rho_{\star\star}[\omega,\sigma] }\,,
\end{flalign}
where $\rho_{\rm true}[\omega,\sigma]$ is the exact smeared spectral density, which is known since the dataset is synthetic.

\begin{figure}[tb]
    \centering
    \includegraphics[width=\globalWidth]{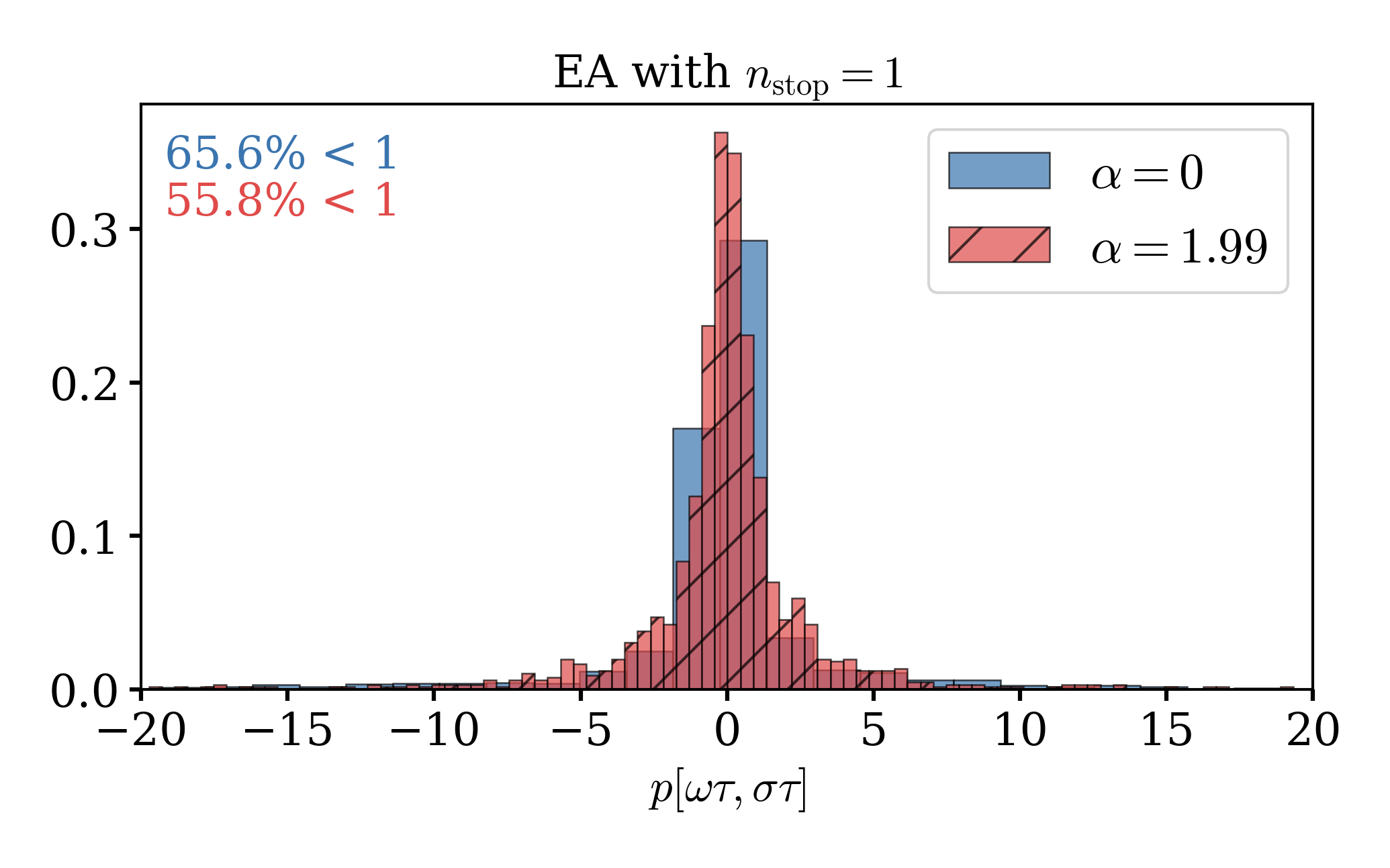}
    \includegraphics[width=\globalWidth]{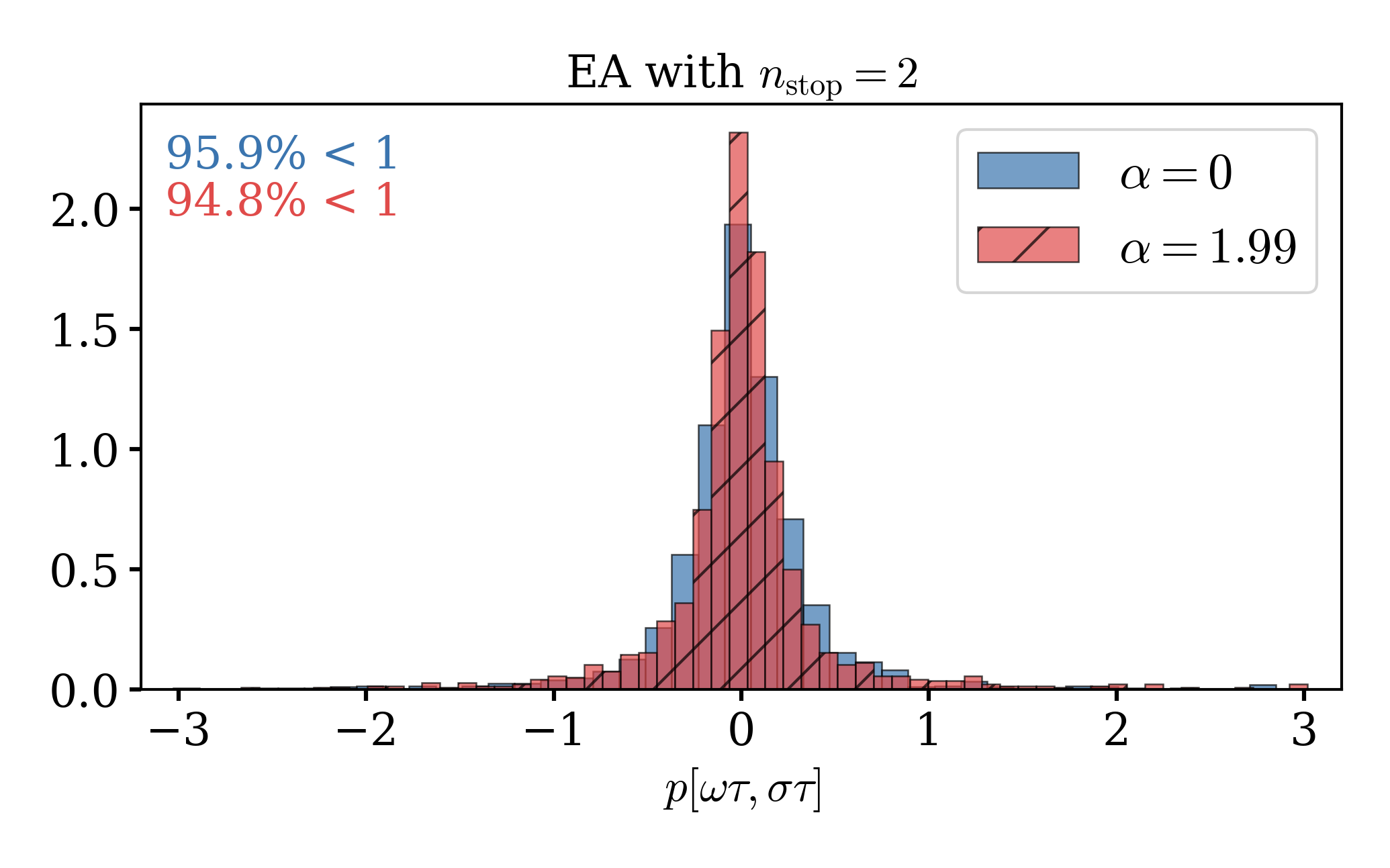}
    \includegraphics[width=\globalWidth]{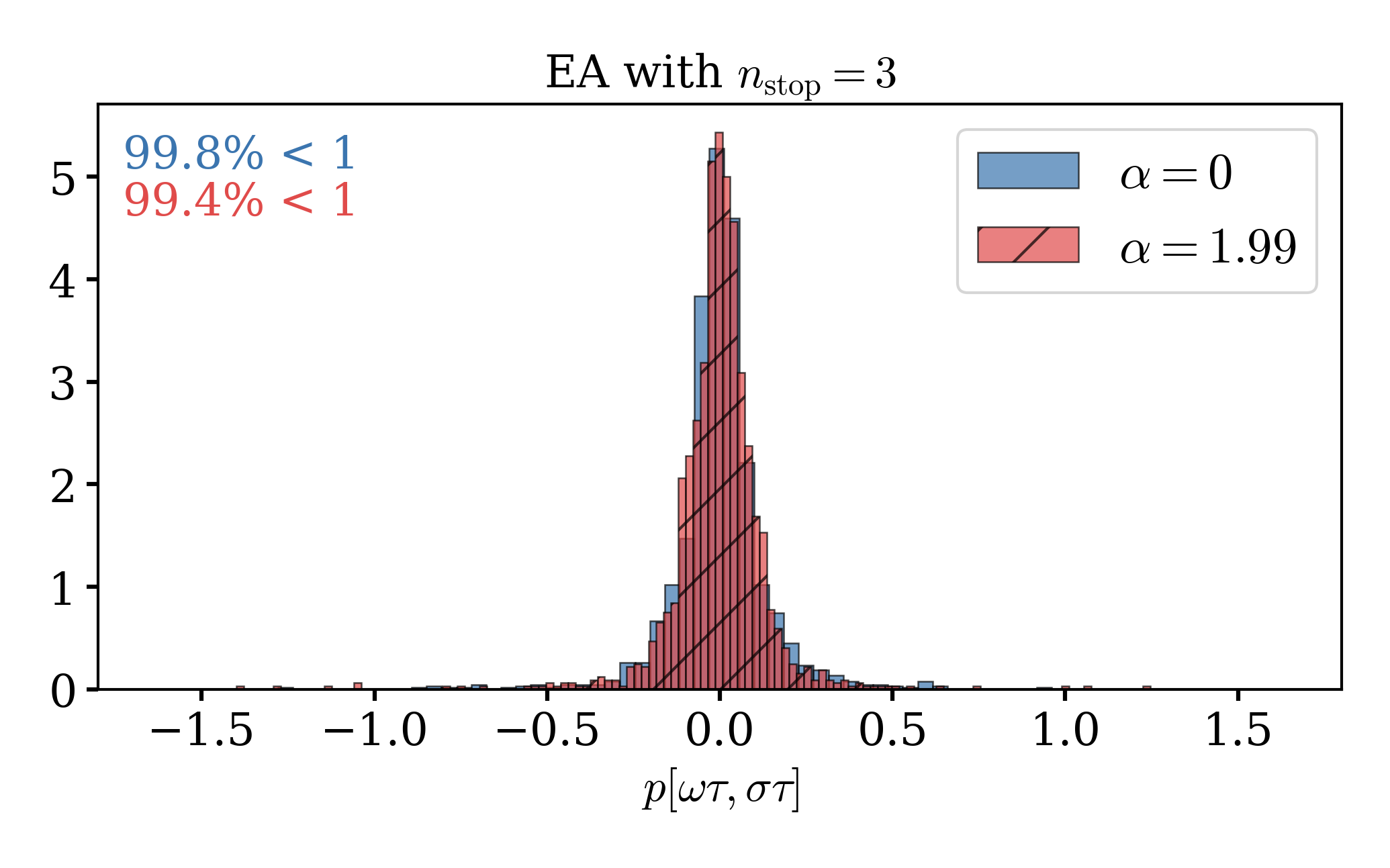}
    \caption{Histogram for the pull defined in Eq.~\eqref{eq:def_pull}, obtained via EA using different values of $\alpha$. Each plot corresponds to a different stopping condition for saturating the sum over the eigenvalues of $\rho[\omega,\sigma]$. The pull shows the reliability of the estimate, while the companion scatter plots, shown in Fig.~\ref{fig:scatter_ET}, suggest whether the quality of the pull distribution comes from a large error estimate rather than a close reproduction of the central values. The parameters of the smeared spectral density are $\omega =  m_{\rho}$,  $\sigma = 2 m_\pi$, while the correlators have $N=64$ points and the statistical noise of a vector-vector lattice QCD correlator.}\label{fig:pull_ET_WTF}
\end{figure}
\begin{figure}[tb]
    \includegraphics[width=\globalWidth]{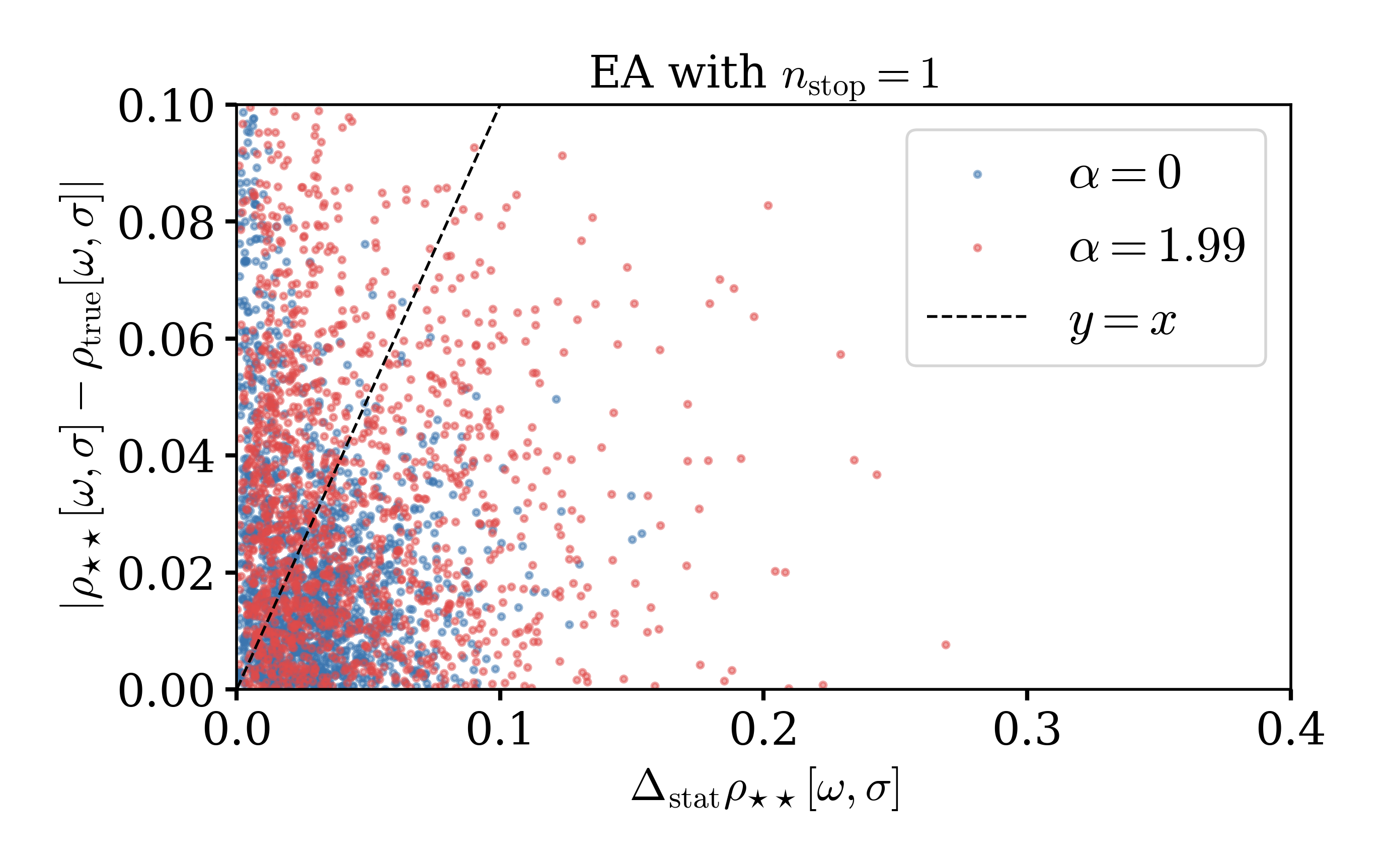}
    \includegraphics[width=\globalWidth]{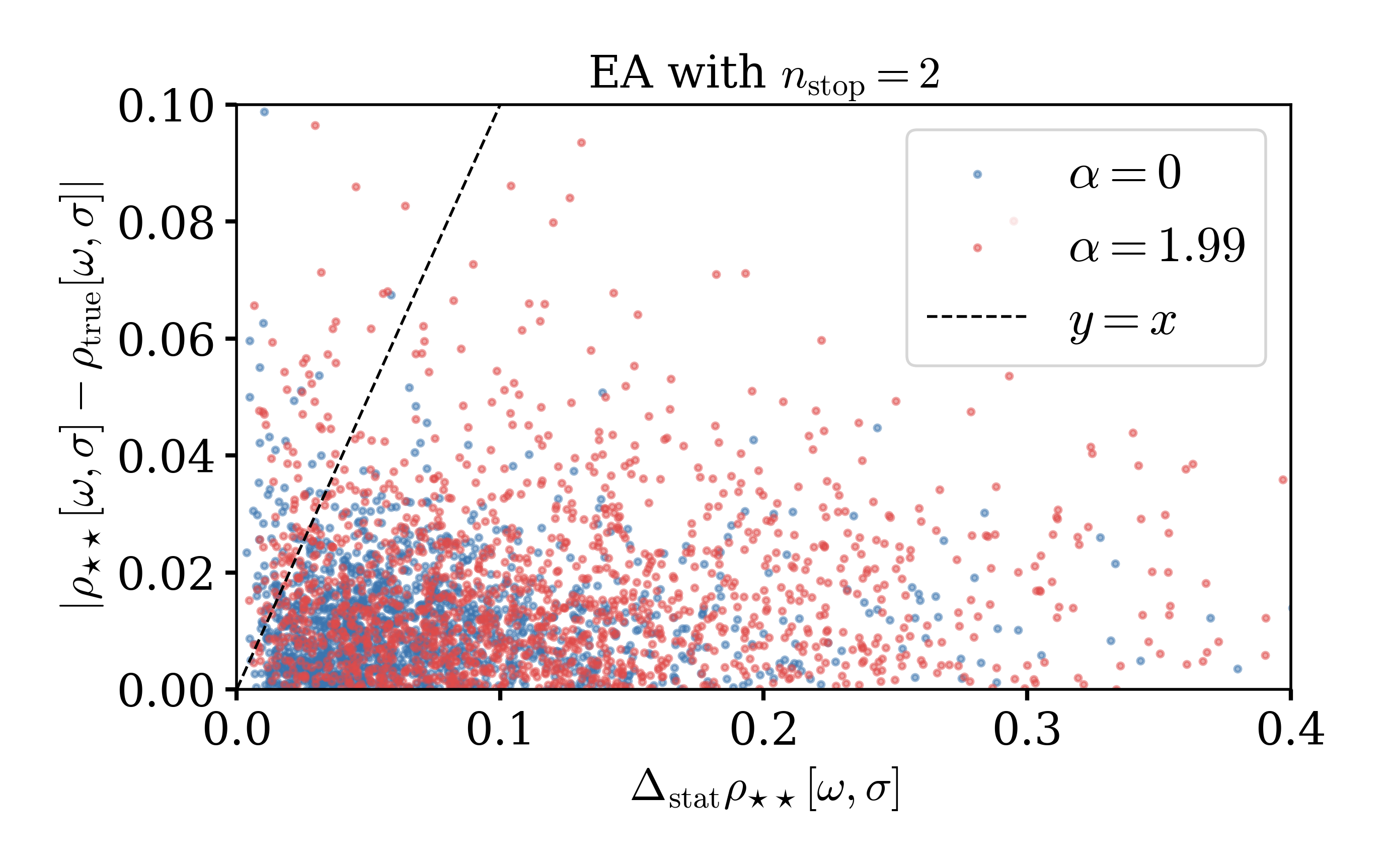}
    \includegraphics[width=\globalWidth]{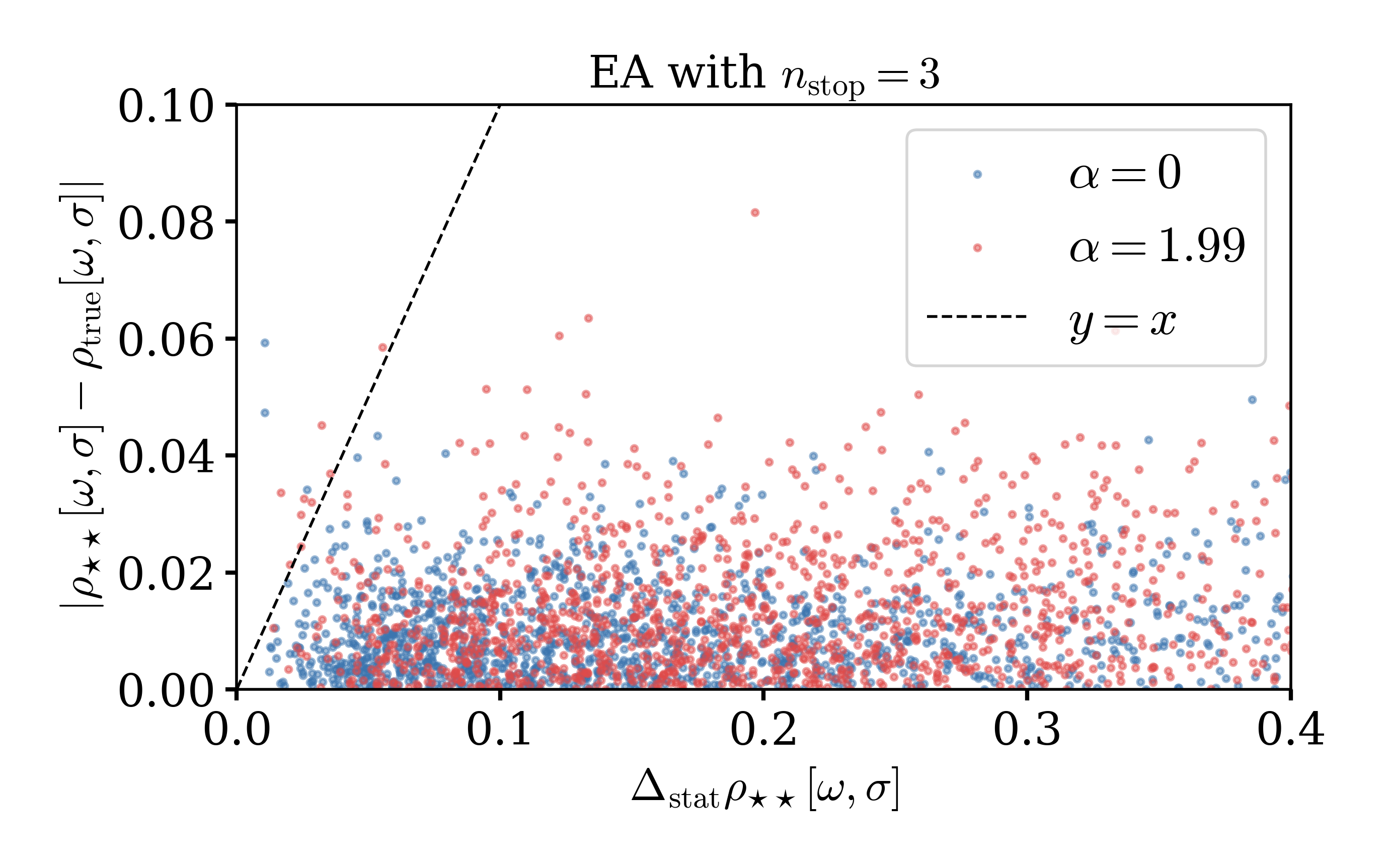}
    \caption{Companion plot of Fig.~\ref{fig:pull_ET_WTF}, showing whether the pull distributions are narrow because of good central values or large errors. The $x$-axis corresponds to the statistical error on the reconstructed smeared spectral density. The $y$-axis corresponds to the modulus of the difference between the reconstructed and the exact smeared spectral density, i.e. the exact systematic error. On the $y=x$ dashed line the systematic and statistical errors are equal.}
    \label{fig:scatter_ET}
\end{figure}

\begin{figure}[tb]
    \centering
    \includegraphics[width=\globalWidth]{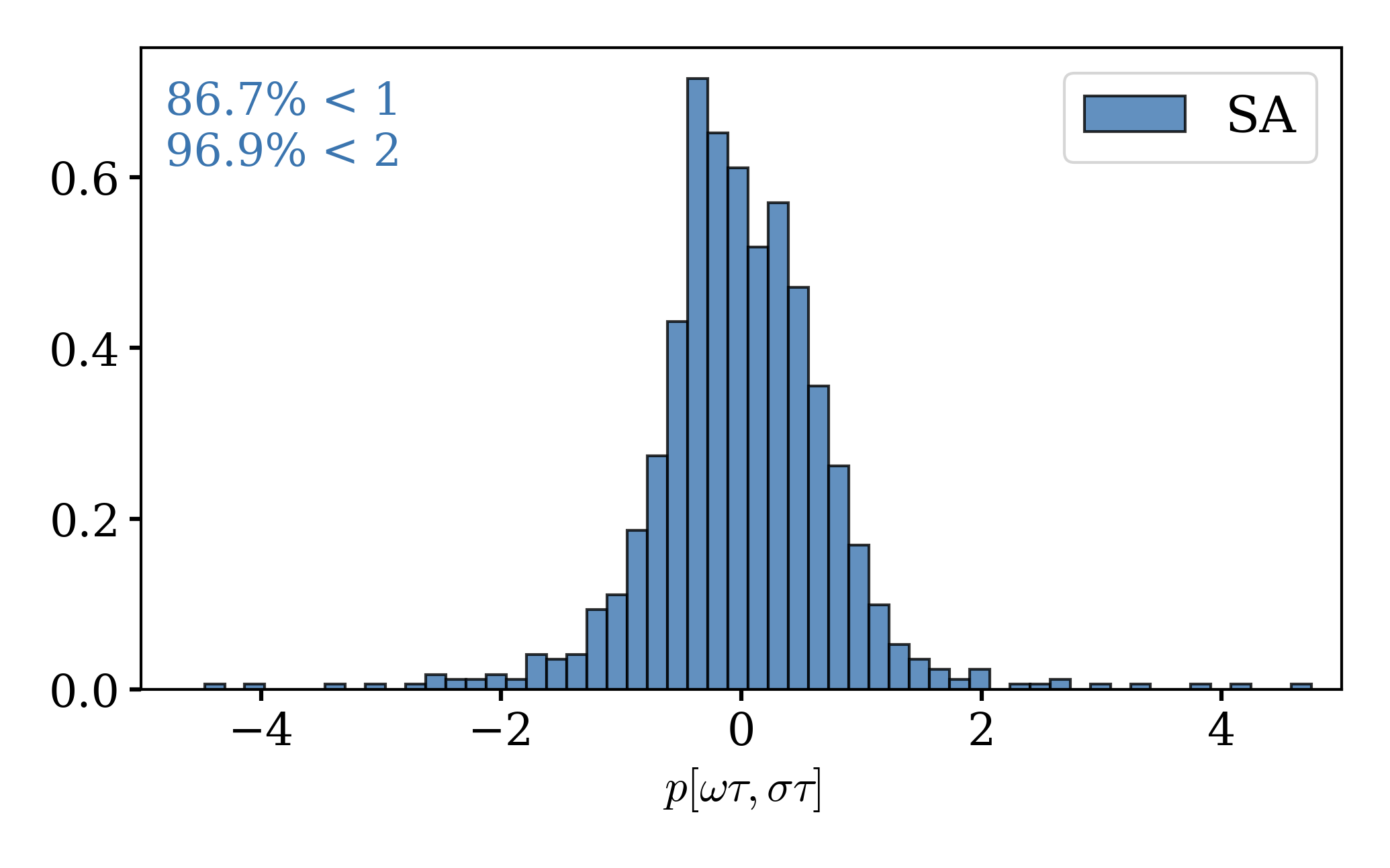}
    \includegraphics[width=\globalWidth]{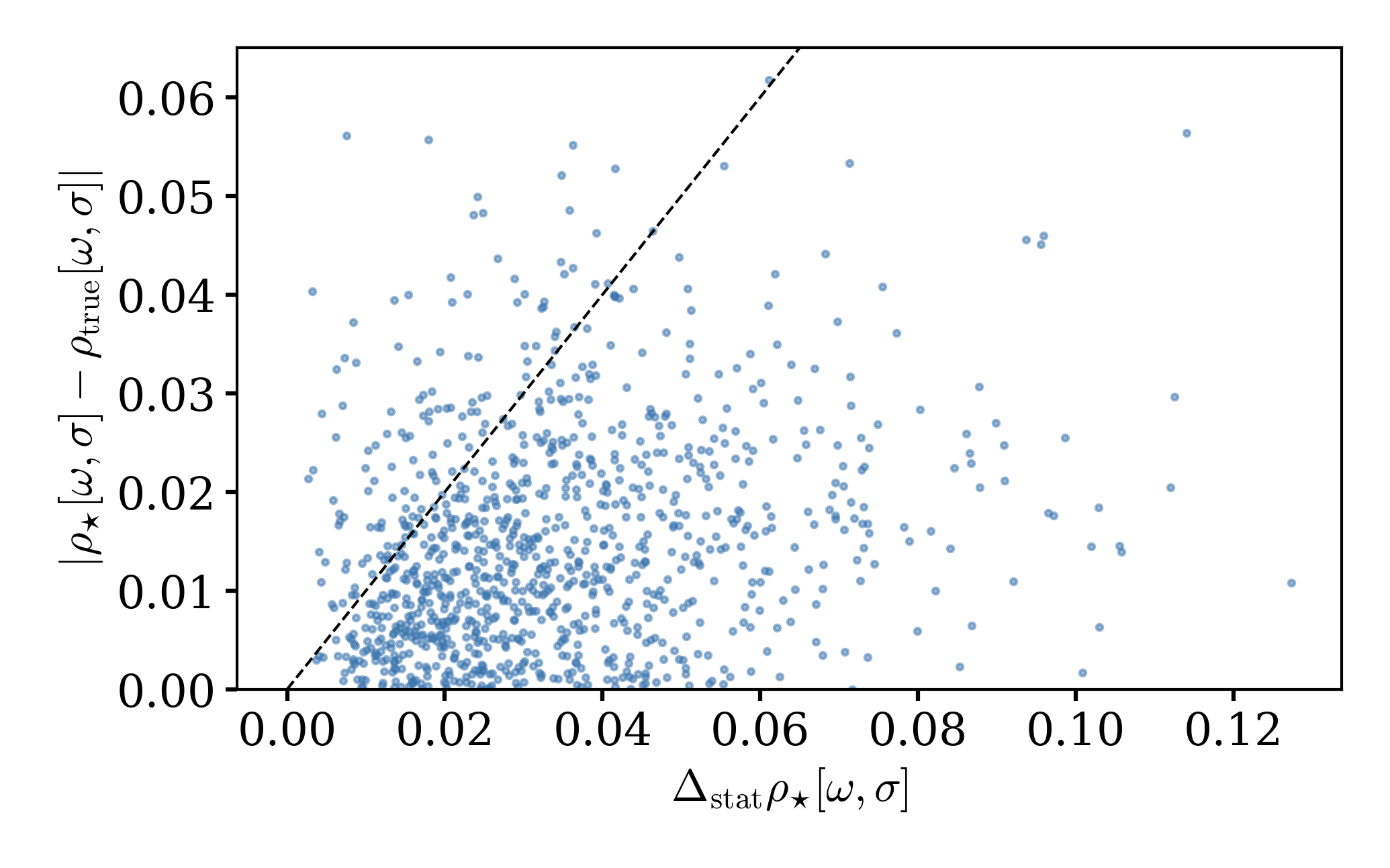}
    \caption{Closure tests on the same datasets, and with the same parameters as Figs.~\ref{fig:pull_ET_WTF} and~\ref{fig:scatter_ET}, this time obtained using the SA to remove the Backus-Gilbert regulator. The results shown here have been obtained by analyzing simultaneously the three values of the norm parameter $\alpha=0, 1, 1.99$.}
    \label{fig:pull_HLT}
\end{figure}
\begin{figure}[tb]
    \centering
    \includegraphics[width=\globalWidth]{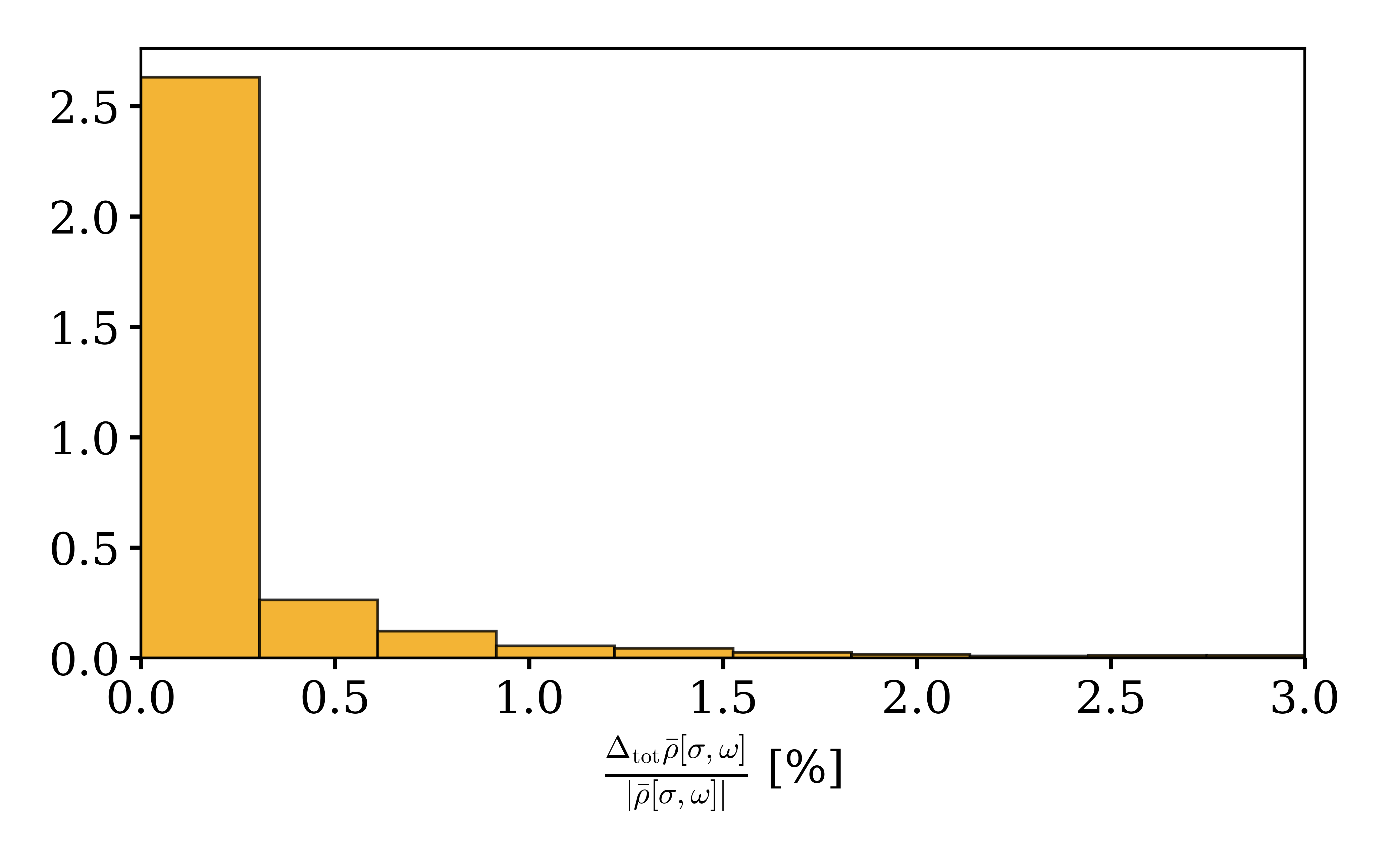}
    \caption{Distribution of the systematic error associated to the difference between the two methods, computed according to Eq.~\eqref{eq:sys_error}, over more than a thousand instances of the inverse problem. Like in the previous examples, $\omega=m_\rho$ and $\sigma=2m_\pi$. The SA uses three values of the norm parameter, $\alpha=0, 1, 1.99$, while the EA uses $n_{\rm stop}=2$.}
    \label{fig:sys_error}
\end{figure}
\begin{figure}[tb]
    \centering
    \includegraphics[width=\globalWidth]{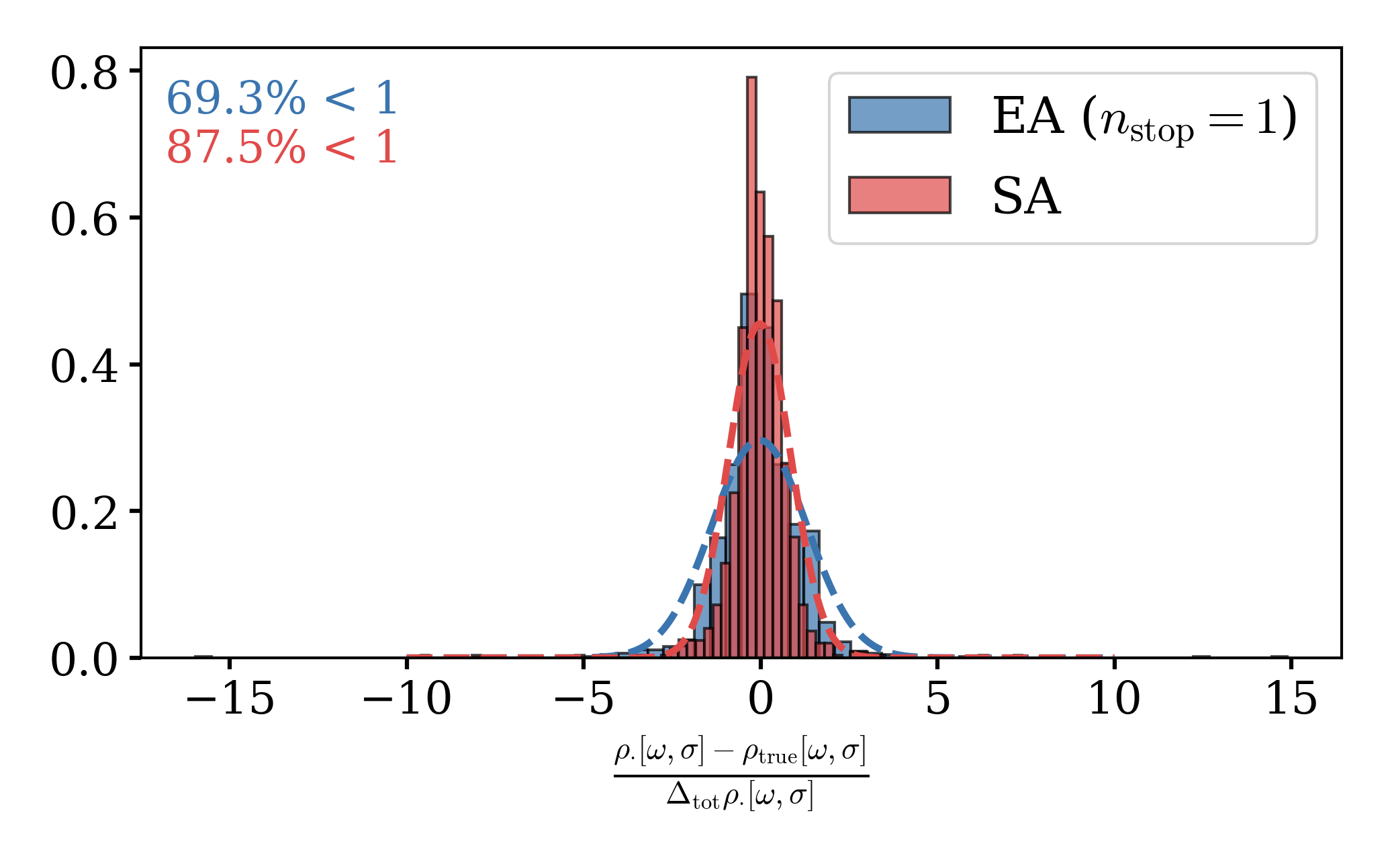}
    \includegraphics[width=\globalWidth]{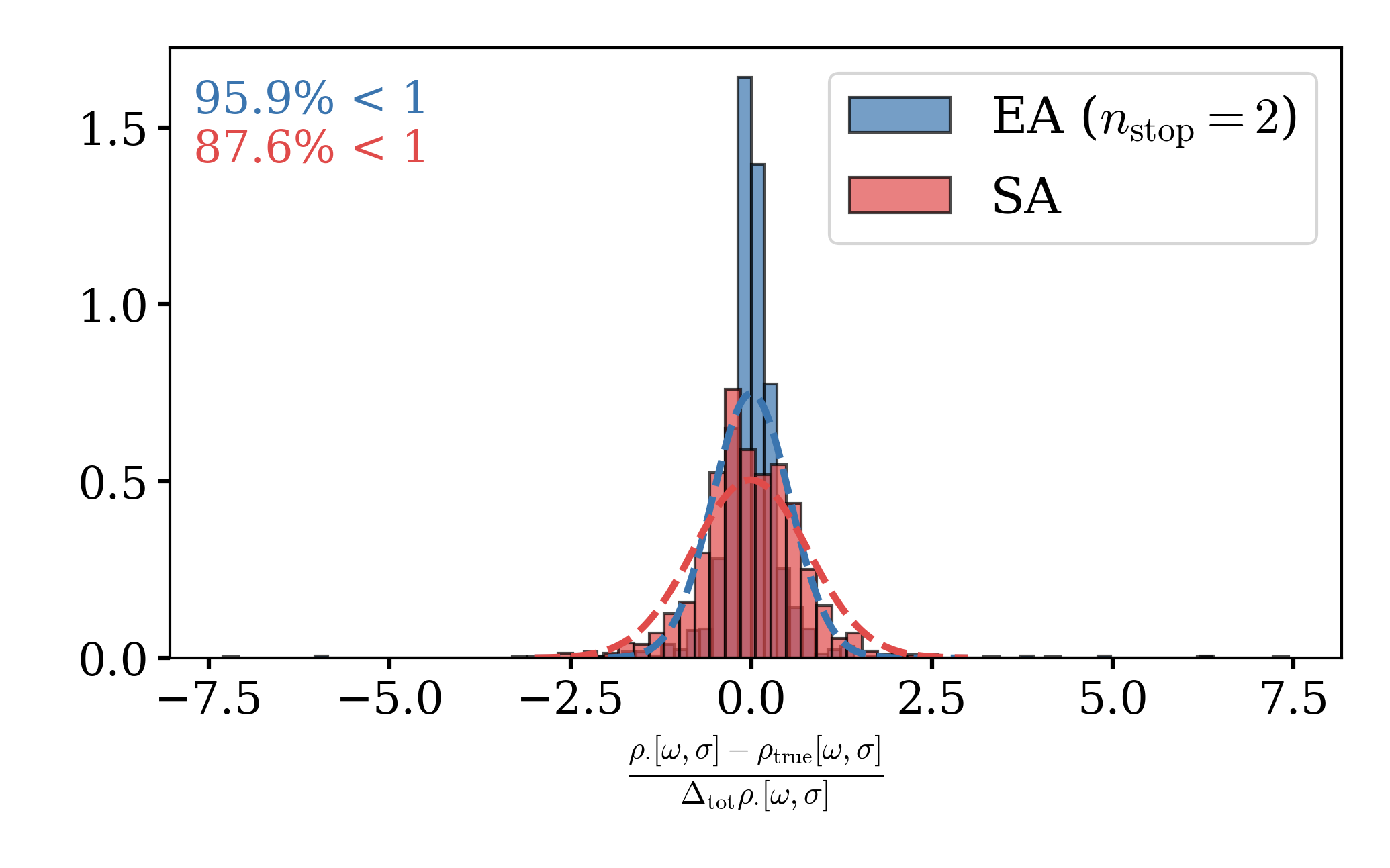}
    \caption{The figure updates Figs.~\ref{fig:pull_ET_WTF} and~\ref{fig:pull_HLT} by considering a pull in which to the statistical error of each regularisation (EA in blue, SA in red) we add in quadrature the systematic error computed according to Eq.~\eqref{eq:sys_error}. This is what in the label of the $x$-axis is denoted with $\Delta_\mathrm{tot}\rho$. In the numerator of the pull, $\rho_{\cdot}$ corresponds to either $\rho_\star$ (SA) or $\rho_{\star\star}$ (EA). For the EA, we show two values of $n_{\rm stop}$ and we use $\alpha=0$. In the SA we used simultaneously $\alpha=0,1,1.99$ \textcolor{MODCOLOR}{(see main text for details)}. The dashed lines are Gaussians defined using the centre and the standard deviation of each histogram.}
    \label{fig:pull_combined}
\end{figure}
\begin{figure}[tb]
    \includegraphics[width=\globalWidth]{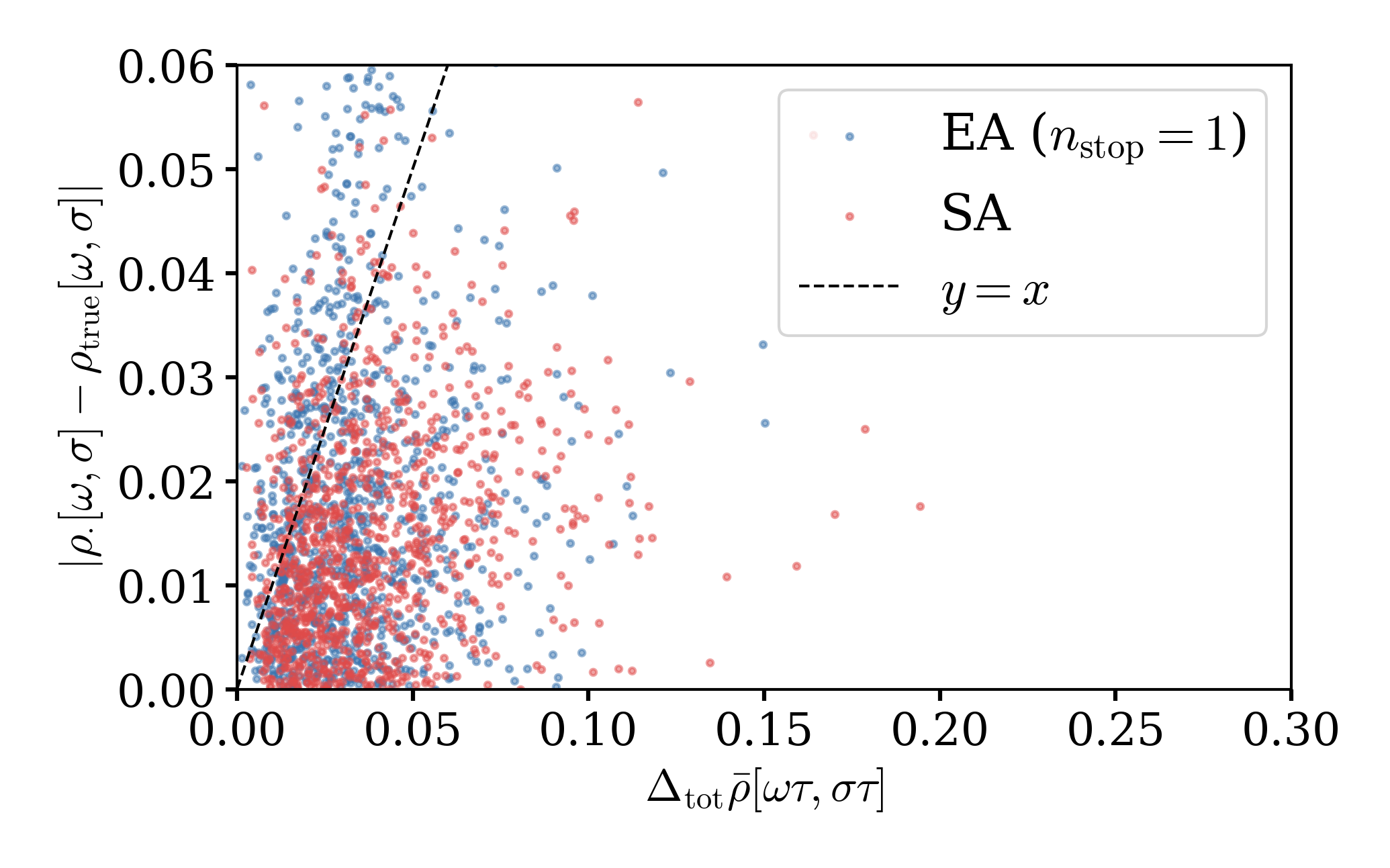}
    \includegraphics[width=\globalWidth]{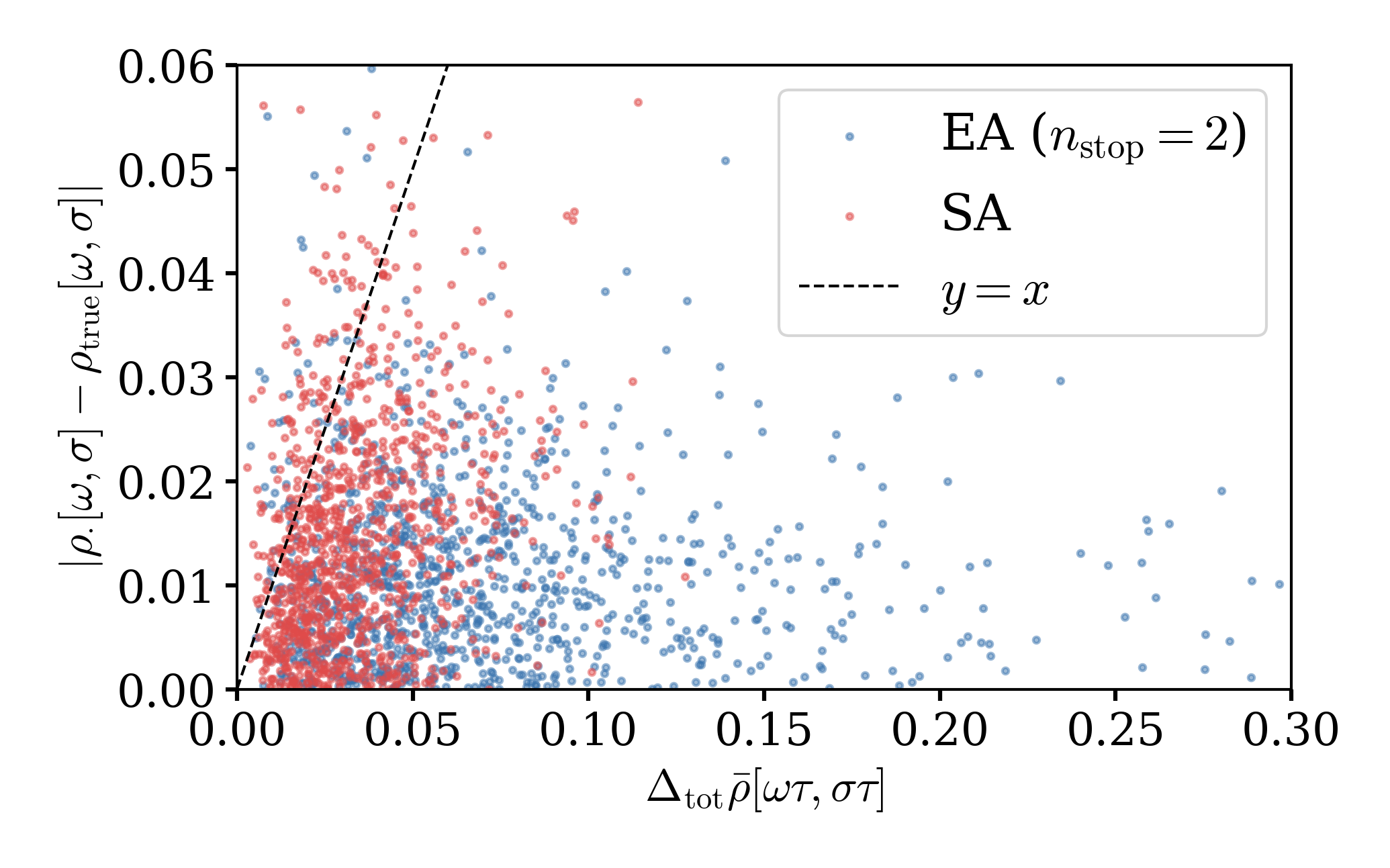}
    \caption{Companion scatter plot of Fig.~\ref{fig:pull_combined} and update of Figs.~\ref{fig:pull_ET_WTF} and~\ref{fig:pull_HLT} with the total error replacing the statistical one.}
    \label{fig:scatter_combined}
\end{figure}

The distribution of such pull variable over the various solutions of the inverse problem is shown in Fig.~\ref{fig:pull_ET_WTF}, for two choices of the norm in Eq.~\eqref{eq:functional_A}, parametrised by two values of $\alpha$. In the same Figure, we display how many estimates of the smeared spectral density are compatible, within one standard deviation, with the true value. We notice that the choice $n_{\rm stop} = 1$, which corresponds to \textcolor{MODCOLOR}{having the sum truncated} as soon as the first term is compatible with zero within its error, is very aggressive and often misses the true value by several standard deviations. On the other hand, by taking $n_{\rm stop} > 1$ one gets very conservative results, with the results corresponding to $n_{\rm stop} = 3$ almost always being compatible with the true values. The scatter plots of Fig.~\ref{fig:scatter_ET} show that, however, the good performance of the $n_{\rm stop} = 3$ procedure is due to the fact that enough terms are included so that the error is already very large. With $\sim 95\%$ of the result compatible with the true value within one standard deviation, the choice $n_{\rm stop} = 2$ appears to be the best option.

In order to compare these performances with the \textcolor{MODCOLOR}{more thoroughly studied ones of the SA}, we display the results on the same input data, using again pulls and associated scatter plots as metrics. The results of the SA, for which we used three values of the norm parameter $\alpha=0, 1, 1.99$ (see Eq.~\eqref{eq:functional_A}) are shown in Fig.~\ref{fig:pull_HLT}. While still being a very conservative procedure, with almost $87\%$ of the results compatible with the true value (notice that only the statistical error is being considered), the SA analysis allows for more aggressive \textcolor{MODCOLOR}{(i.e. with smaller errors)} results on average.

We now wish to ask whether the two methods can be used in conjunction, in order to assess whether the EA can help identify biases in the SA analysis and vice versa. To this end, we take
\begin{flalign}
\bar \rho[\mathcal S] = \frac{\rho_\star[\mathcal S] +\rho_{\star\star}[\mathcal S]}{2}\,,
    \label{eq:central}
\end{flalign}
as the new estimator of $\rho[\mathcal S]$ and introduce an estimator of the systematic error by using the difference between the two methods, according to
\begin{flalign}
    \Delta_{\rm sys}\bar \rho[\mathcal S] = \left\vert \rho_\star[\mathcal S] -\rho_{\star\star}[\mathcal S]\right\vert 
    \text{Erf} \left( \frac{\rho_\star[\mathcal S] -\rho_{\star\star}[\mathcal S]}{\sqrt{2}\Delta_\mathrm{stat} \bar \rho[\mathcal S]} \right) \, .
    \label{eq:sys_error}
\end{flalign}
The modulation with the error-function is meant to account for the possibility that the difference is due to statistical fluctuations. The distribution of this systematic difference over the closure tests is shown in Fig.~\ref{fig:sys_error}. In most cases, the difference between the two approaches lies below $1\%$ of the central value, raising occasionally to a few percentages.
Performances including this additional systematics are shown in Fig.~\ref{fig:pull_combined} in terms of the pull distribution, and the associated scatter plot in Fig.~\ref{fig:scatter_combined}. \textcolor{MODCOLOR}{To produce this plot, we used three values of $\alpha$ in the SA, as described in Section~\ref{sec:stability}, and a single value of $\alpha$ for the EA, since different values produced the same results}. As expected, the difference is smaller in the cases where one of the two approaches is already very conservative, but plays a role when the error is possibly underestimated. This is the case of the EA with $n_{\rm stop}=1$  where the situation substantially improves by adding this systematic error.

Motivated by these results, our suggestion is to combine the two procedures, by using $n_{\rm stop}=2$ to define $\rho_{\star\star}[\mathcal S]$ and Eq.~\eqref{eq:sys_error} to define the systematic error on the central value, i.e.\ $\bar \rho[\mathcal S]$ defined in Eq.~\eqref{eq:central}, so that the total error is given by
\begin{flalign}
    \Delta_{\rm tot}\bar \rho[\mathcal S] = \sqrt{
    \left( \Delta_{\rm stat}\bar \rho[\mathcal S]\right)^2
    +
    \left( \Delta_{\rm sys}\bar \rho[\mathcal S]\right)^2
    }\, .
    \label{eq:tot_error}
\end{flalign}

%% file: sections/conclusions.tex
\section{Conclusion}
\label{sec:conclusions}

We presented an alternative implementation of the HLT method~\cite{Hansen:2019idp} which does not necessarily rely on a Backus-Gilbert regularisation. This is possible because the smeared spectral density, which is obtained as a linear combination of Euclidean correlators, displays a peculiar behaviour after a change of basis: the terms contributing the most to the signal are the most precise, and as the sum saturates, the error begins to increase. This leaves the possibility for the existence of a window in which the sum can be truncated, with the effect of limiting the error without affecting the signal. We refer to this procedure as eigen-space analysis.

In practice, for state-of-the-art data and parameters of the reconstructions, such window counts two or three terms. The options for truncating the sum are very limited. On one hand, this leaves little freedom: our tests suggest that the truncation produces errors which are either very aggressive or very conservative. Such conclusions are derived from closure tests that we performed, in this work, in order to test the performance of this method in the realistic situation of a state-of-the-art lattice simulation. On the other hand, it is worth stressing the simplicity of this new procedure. Compared to the original HLT stability analysis, the eigen-space analysis requires much less tuning and its automation is very straightforward. This method can be therefore useful for a quick estimate of the solution.

Our test suggests that a reasonable way to cut the eigen-space sum is to stop including the second term that is compatible with zero. This and other choices were tested in comparison with the traditional stability analysis that has been used to {\color{MODCOLOR} numerically} remove the bias due to the Backus-Gilbert regulator. The latter seems preferable, since the eigen-space truncation, on average, tends to overestimate the errors more often.
In order to benefit from both analyses, we proposed in Section~\ref{sec:pulls} a way to combine the two. We have shown how accounting for the systematic difference between the two approaches can improve the reliability of the results.

%% file: sections/appendix.tex
\section{On the order of the $a\to 0$, $L\to \infty$, $N\to \infty$, $\lambda\to 0$ limits.}
\label{sec:appendix}

The reconstruction of the physical spectral density from lattice simulations requires the introduction of a number of intermediate regulators, some inherent to the lattice (the lattice spacing $a$ and the finite volume $L$), some to the resolution of the inverse problem ($N$ and possibly $\lambda$). 

At fixed $a$ and $L$ a generic spectral density has the form
\begin{flalign}
\rho(E;a,L)=\sum_{n} w_n(a,L)\, \delta\left(E-E_n(a,L) \right)\,,
\end{flalign}
where the sum runs on the discrete eigenvalues $E_n(a,L)$ of the lattice finite-volume hamiltonian and $w_n(a,L)$ are the associated spectral weights. These are on-shell quantities.  

The spectral density $\rho(E;a,L)$ is a distribution and, therefore, taking its continuum and/or infinite-volume limit is not a well defined operation (at least from the computational perspective). On the contrary, the $a\to 0$ and $L\to \infty$ limits of the smeared spectral density,
\begin{flalign} 
\rho[\mathcal S;a,L] = \int_{E_0}^\infty dE\, \mathcal{S}(E)\, \rho(E;a,L)\,,
\end{flalign}
are perfectly well defined, provided that $E_0$ and $\mathcal{S}(E)$ are \emph{kept fixed}. 
The necessity of introducing a smearing kernel to properly define the infinite-volume limit has been extensively discussed in the previous literature on the subject, see e.g.\ \cite{Hansen:2017mnd, Hansen:2019idp, Bulava:2021fre}.  

In the HLT approximation strategy two approaches are possible for taking the continuum and infinite-volume limits. To explain these approaches we introduce the lattice correlator (see Eq.~\eqref{eq:laplace_transform}),
\begin{flalign}
    C(n \tau;a,L) = \int_{E_0}^\infty dE \; e^{-n\tau E} \,\rho(E;a,L) \, ,
    \label{eq:laplace_transform_a}
\end{flalign}
which, see the main text, allows us to write
\begin{flalign} 
\rho_{\scriptstyle N,\lambda}[\mathcal{S};a,L] = 
\sum_{n=1}^N g_{\scriptstyle [\mathcal{S}],N,\lambda}(n)\, C(n \tau;a,L) \,.
\label{eq:discrete_series}
\end{flalign}

The first option consists in taking $\tau$ fixed in physical units, e.g.\ $\tau=0.1$~fm, and considering the following order of the limits
\begin{flalign} 
&
C(n \tau) =\lim_{a\to 0, L\to \infty} C(n \tau;a,L)\,,
\nonumber \\[8pt]
&
\rho[\mathcal{S}]
=
\lim_{\lambda\to 0, N\to \infty}
\sum_{n=1}^N g_{\scriptstyle [\mathcal{S}],N,\lambda}(n)\, C(n \tau) \,,
\end{flalign}
i.e.\ the lattice regulator is removed by extrapolating the lattice correlator at fixed $\tau$ and before the algorithmic limits are performed. This requires a robust procedure to interpolate the lattice data $C(n a;a,L)$ in order to get the correlator at the times $n\tau$, i.e.\ $C(n \tau;a,L)$. Such a step introduces an additional systematic error associated with the interpolation.  This approach has been studied in mathematical details in Ref.~\cite{Patella:2024cto}, where the original work~\cite{Barata:1990rn} of Barata and Fredenhagen has been generalised to this case.

The second option, the one originally considered by Barata and Fredenhagen and used in previous applications~\cite{Bulava:2021fre,ExtendedTwistedMassCollaborationETMC:2022sta, ExtendedTwistedMass:2024myu, Evangelista:2023fmt, DeSantis:2025qbb, DeSantis:2025yfm, Bonanno:2023ljc, Bonanno:2023thi}, consists in identifying $\tau$ with the lattice spacing, i.e.\ in setting $\tau=a$, and considering the following order of the limits
\begin{flalign} 
&
\rho[\mathcal{S};a,L]
=
\lim_{\lambda\to 0, N\to \infty}
\rho_{\scriptstyle N,\lambda}[\mathcal{S};a,L]\,,
\nonumber \\[8pt]
&
\rho[\mathcal{S}]
=
\lim_{a\to 0, L\to \infty}
\rho[\mathcal{S};a,L]\,.
\end{flalign}
In this second approach it is crucially important that the algorithmic limits are performed \emph{before} the removal of the lattice regulator. This ensures that one has the \emph{same} smearing kernel at all values of $a$ and $L$ and, therefore, that the continuum limit is perfectly well defined. 

The fact that both approaches are possible\footnote{The approach of Ref.~\cite{Bruno:2024fqc} {\color{MODCOLOR} considers smearing kernels} that can be represented as $\mathcal{S}(E)=\int_0^\infty dt\, g(t) e^{-tE}$, i.e.\ as integrals that correspond to the ``continuum limit'' of the discrete series of Eq.~(\ref{eq:discrete_series}). This is in fact a third option for taking the continuum limit of smeared spectral densities reconstructed from lattice correlators and we refer to~\cite{Bruno:2024fqc} for further details.}, and equally legitimate, is a direct consequence of the Stone-Weierstrass theorem and, in the main body of this paper, we have discussed how the limits $N\rightarrow \infty$ and $\lambda\to 0$ can be taken by means of the stability analysis. If the errors (both statistical and systematics) associated with the stability analysis have been properly estimated, the algorithmic $N\rightarrow \infty$ and $\lambda\to 0$ limits have been taken and, within the uncertainties, one has the same smearing kernel $\mathcal{S}(E)$ at all the simulated lattice spacings and volumes. This is the reason why the main subject of this paper, i.e.\ a procedure to get a robust estimate of these errors, is a crucial matter.







%% file: main.bib
@article{DelDebbio:2022qgu,
    author = "Del Debbio, Luigi and Lupo, Alessandro and Panero, Marco and Tantalo, Nazario",
    title = "{Multi-representation dynamics of SU(4) composite Higgs models: chiral limit and spectral reconstructions}",
    eprint = "2211.09581",
    archivePrefix = "arXiv",
    primaryClass = "hep-lat",
    doi = "10.1140/epjc/s10052-023-11363-8",
    journal = "Eur. Phys. J. C",
    volume = "83",
    number = "3",
    pages = "220",
    year = "2023"
}

@article{Barata:1990rn,
    author = "Barata, J. C. A. and Fredenhagen, K.",
    title = "{Particle scattering in Euclidean lattice field theories}",
    doi = "10.1007/BF02102039",
    journal = "Commun. Math. Phys.",
    volume = "138",
    pages = "507--520",
    year = "1991"
}

@article{Furmanski:1981ja,
    author = "Furmanski, W. and Petronzio, R.",
    title = "{A Method of Analyzing the Scaling Violation of Inclusive Spectra in Hard Processes}",
    reportNumber = "CERN-TH-3047",
    doi = "10.1016/0550-3213(82)90398-4",
    journal = "Nucl. Phys. B",
    volume = "195",
    pages = "237--261",
    year = "1982"
}

@article{Hansen:2017mnd,
    author = "Hansen, Maxwell T. and Meyer, Harvey B. and Robaina, Daniel",
    title = "{From deep inelastic scattering to heavy-flavor semileptonic decays: Total rates into multihadron final states from lattice QCD}",
    eprint = "1704.08993",
    archivePrefix = "arXiv",
    primaryClass = "hep-lat",
    doi = "10.1103/PhysRevD.96.094513",
    journal = "Phys. Rev. D",
    volume = "96",
    number = "9",
    pages = "094513",
    year = "2017"
}

@article{Bulava:2021fre,
    author = "Bulava, John and Hansen, Maxwell T. and Hansen, Michael W. and Patella, Agostino and Tantalo, Nazario",
    title = "{Inclusive rates from smeared spectral densities in the two-dimensional O(3) non-linear \ensuremath{\sigma}-model}",
    eprint = "2111.12774",
    archivePrefix = "arXiv",
    primaryClass = "hep-lat",
    reportNumber = "DESY 21-201, HU-EP-21/49",
    doi = "10.1007/JHEP07(2022)034",
    journal = "JHEP",
    volume = "07",
    pages = "034",
    year = "2022"
}

@article{DelDebbio:2024lwm,
    author = "Del Debbio, Luigi and Lupo, Alessandro and Panero, Marco and Tantalo, Nazario",
    title = "{Bayesian solution to the inverse problem and its relation to Backus{\textendash}Gilbert methods}",
    eprint = "2409.04413",
    archivePrefix = "arXiv",
    primaryClass = "hep-lat",
    doi = "10.1140/epjc/s10052-025-13885-9",
    journal = "Eur. Phys. J. C",
    volume = "85",
    number = "2",
    pages = "185",
    year = "2025"
}

@article{10.1111/j.1365-246X.1968.tb00216.x,
    author = {Backus, George and Gilbert, Freeman},
    title = "{The Resolving Power of Gross Earth Data}",
    journal = {Geophysical Journal International},
    volume = {16},
    number = {2},
    pages = {169-205},
    year = {1968},
    month = {10},
    issn = {0956-540X},
    doi = {10.1111/j.1365-246X.1968.tb00216.x},
    url = {https://doi.org/10.1111/j.1365-246X.1968.tb00216.x},
}

@article{Bruno:2024fqc,
    author = "Bruno, Mattia and Giusti, Leonardo and Saccardi, Matteo",
    title = "{Spectral densities from Euclidean lattice correlators via the Mellin transform}",
    eprint = "2407.04141",
    archivePrefix = "arXiv",
    primaryClass = "hep-lat",
    doi = "10.1103/PhysRevD.111.094515",
    journal = "Phys. Rev. D",
    volume = "111",
    number = "9",
    pages = "094515",
    year = "2025"
}

@article{Hansen:2019idp,
    author = "Hansen, Martin and Lupo, Alessandro and Tantalo, Nazario",
    title = "{Extraction of spectral densities from lattice correlators}",
    eprint = "1903.06476",
    archivePrefix = "arXiv",
    primaryClass = "hep-lat",
    doi = "10.1103/PhysRevD.99.094508",
    journal = "Phys. Rev. D",
    volume = "99",
    number = "9",
    pages = "094508",
    year = "2019"
}

@article{Buzzicotti:2023qdv,
    author = "Buzzicotti, Michele and De Santis, Alessandro and Tantalo, Nazario",
    title = "{Teaching to extract spectral densities from lattice correlators to a broad audience of learning-machines}",
    eprint = "2307.00808",
    archivePrefix = "arXiv",
    primaryClass = "hep-lat",
    doi = "10.1140/epjc/s10052-024-12399-0",
    journal = "Eur. Phys. J. C",
    volume = "84",
    number = "1",
    pages = "32",
    year = "2024"
}

@article{DelDebbio:2021whr,
    author = "Del Debbio, Luigi and Giani, Tommaso and Wilson, Michael",
    title = "{Bayesian approach to inverse problems: an application to NNPDF closure testing}",
    eprint = "2111.05787",
    archivePrefix = "arXiv",
    primaryClass = "hep-ph",
    doi = "10.1140/epjc/s10052-022-10297-x",
    journal = "Eur. Phys. J. C",
    volume = "82",
    number = "4",
    pages = "330",
    year = "2022"
}

@article{ExtendedTwistedMassCollaborationETMC:2022sta,
    author = "Alexandrou, Constantia and others",
    collaboration = "Extended Twisted Mass",
    title = "{Probing the Energy-Smeared R Ratio Using Lattice QCD}",
    eprint = "2212.08467",
    archivePrefix = "arXiv",
    primaryClass = "hep-lat",
    doi = "10.1103/PhysRevLett.130.241901",
    journal = "Phys. Rev. Lett.",
    volume = "130",
    number = "24",
    pages = "241901",
    year = "2023"
}

@article{Bergamaschi:2023xzx,
    author = "Bergamaschi, Thomas and Jay, William I. and Oare, Patrick R.",
    title = "{Hadronic structure, conformal maps, and analytic continuation}",
    eprint = "2305.16190",
    archivePrefix = "arXiv",
    primaryClass = "hep-lat",
    reportNumber = "MIT-CTP/5563",
    doi = "10.1103/PhysRevD.108.074516",
    journal = "Phys. Rev. D",
    volume = "108",
    number = "7",
    pages = "074516",
    year = "2023"
}

@article{Bonanno:2023thi,
    author = "Bonanno, Claudio and D'Angelo, Francesco and D'Elia, Massimo and Maio, Lorenzo and Naviglio, Manuel",
    title = "{Sphaleron Rate of Nf=2+1 QCD}",
    eprint = "2308.01287",
    archivePrefix = "arXiv",
    primaryClass = "hep-lat",
    doi = "10.1103/PhysRevLett.132.051903",
    journal = "Phys. Rev. Lett.",
    volume = "132",
    number = "5",
    pages = "051903",
    year = "2024"
}

@article{Barone:2023tbl,
    author = {Barone, Alessandro and Hashimoto, Shoji and J\"uttner, Andreas and Kaneko, Takashi and Kellermann, Ryan},
    title = "{Approaches to inclusive semileptonic B$_{(s)}$-meson decays from Lattice QCD}",
    eprint = "2305.14092",
    archivePrefix = "arXiv",
    primaryClass = "hep-lat",
    reportNumber = "KEK-CP-0394 CERN-TH-2023-087",
    doi = "10.1007/JHEP07(2023)145",
    journal = "JHEP",
    volume = "07",
    pages = "145",
    year = "2023"
}

@article{Bennett:2024cqv,
    author = "Bennett, Ed and others",
    title = "{Meson spectroscopy from spectral densities in lattice gauge theories}",
    eprint = "2405.01388",
    archivePrefix = "arXiv",
    primaryClass = "hep-lat",
    reportNumber = "CTPU-PTC-24-11, PNUTP-24/A02",
    doi = "10.1103/PhysRevD.110.074509",
    journal = "Phys. Rev. D",
    volume = "110",
    number = "7",
    pages = "074509",
    year = "2024"
}

@article{Patella:2024cto,
    author = "Patella, Agostino and Tantalo, Nazario",
    title = "{Scattering amplitudes from Euclidean correlators: Haag-Ruelle theory and approximation formulae}",
    eprint = "2407.02069",
    archivePrefix = "arXiv",
    primaryClass = "hep-lat",
    doi = "10.1007/JHEP01(2025)091",
    journal = "JHEP",
    volume = "01",
    pages = "091",
    year = "2025"
}

@article{Burnier:2013nla,
    author = "Burnier, Yannis and Rothkopf, Alexander",
    title = "{Bayesian Approach to Spectral Function Reconstruction for Euclidean Quantum Field Theories}",
    eprint = "1307.6106",
    archivePrefix = "arXiv",
    primaryClass = "hep-lat",
    doi = "10.1103/PhysRevLett.111.182003",
    journal = "Phys. Rev. Lett.",
    volume = "111",
    pages = "182003",
    year = "2013"
}

@article{Medrano:2025cmg,
    author = "Medrano, Yamil Cahuana and Dutrieux, Herv{\'e} and Karpie, Joseph and Orginos, Kostas and Zafeiropoulos, Savvas",
    title = "{Gaussian Processes for Inferring Parton Distributions}",
    eprint = "2510.21041",
    archivePrefix = "arXiv",
    primaryClass = "hep-lat",
    reportNumber = "JLAB-THY-25-4579",
    month = "10",
    year = "2025"
}

@article{DelDebbio:2020rgv,
    author = "Del Debbio, Luigi and Giani, Tommaso and Karpie, Joseph and Orginos, Kostas and Radyushkin, Anatoly and Zafeiropoulos, Savvas",
    title = "{Neural-network analysis of Parton Distribution Functions from Ioffe-time pseudodistributions}",
    eprint = "2010.03996",
    archivePrefix = "arXiv",
    primaryClass = "hep-ph",
    doi = "10.1007/JHEP02(2021)138",
    journal = "JHEP",
    volume = "02",
    pages = "138",
    year = "2021"
}

@article{Bonanno:2023ljc,
    author = "Bonanno, Claudio and D'Angelo, Francesco and D'Elia, Massimo and Maio, Lorenzo and Naviglio, Manuel",
    title = "{Sphaleron rate from a modified Backus-Gilbert inversion method}",
    eprint = "2305.17120",
    archivePrefix = "arXiv",
    primaryClass = "hep-lat",
    doi = "10.1103/PhysRevD.108.074515",
    journal = "Phys. Rev. D",
    volume = "108",
    number = "7",
    pages = "074515",
    year = "2023"
}

@article{pijpers1994sola,
  title={The SOLA method for helioseismic inversion},
  author={Pijpers, Frank P and Thompson, Michael John},
  journal={Astronomy and Astrophysics (ISSN 0004-6361), vol. 281, no. 1, p. 231-240},
  volume={281},
  pages={231--240},
  year={1994}
}

@article{Bertero_1988,
doi = {10.1088/0266-5611/4/3/004},
url = {https://doi.org/10.1088/0266-5611/4/3/004},
year = {1988},
month = {aug},
publisher = {},
volume = {4},
number = {3},
pages = {573},
author = {M Bertero and C De Mol and E R Pike},
title = {Linear inverse problems with discrete data: {II}. Stability and regularisation},
journal = {Inverse Problems}
}

@article{DeSantis:2025yfm,
    author = "De Santis, Alessandro and others",
    title = "{Inclusive Semileptonic Decays of the Ds Meson: Lattice QCD Confronts Experiments}",
    eprint = "2504.06064",
    archivePrefix = "arXiv",
    primaryClass = "hep-lat",
    reportNumber = "HIP-2025-10/TH",
    doi = "10.1103/snc6-cpz6",
    journal = "Phys. Rev. Lett.",
    volume = "135",
    number = "12",
    pages = "121901",
    year = "2025"
}

@article{DeSantis:2025qbb,
    author = "De Santis, Alessandro and others",
    title = "{Inclusive semileptonic decays of the Ds meson: A first-principles lattice QCD calculation}",
    eprint = "2504.06063",
    archivePrefix = "arXiv",
    primaryClass = "hep-lat",
    reportNumber = "HIP-2025-9/TH",
    doi = "10.1103/3cxg-k322",
    journal = "Phys. Rev. D",
    volume = "112",
    number = "5",
    pages = "054503",
    year = "2025"
}

@article{ExtendedTwistedMass:2024myu,
    author = "Alexandrou, Constantia and others",
    collaboration = "Extended Twisted Mass",
    title = "{Inclusive Hadronic Decay Rate of the {\ensuremath{\tau}} Lepton from Lattice QCD: The u{\textasciimacron}s Flavor Channel and the Cabibbo Angle}",
    eprint = "2403.05404",
    archivePrefix = "arXiv",
    primaryClass = "hep-lat",
    doi = "10.1103/PhysRevLett.132.261901",
    journal = "Phys. Rev. Lett.",
    volume = "132",
    number = "26",
    pages = "261901",
    year = "2024"
}

@article{Evangelista:2023fmt,
    author = "Evangelista, Antonio and Frezzotti, Roberto and Tantalo, Nazario and Gagliardi, Giuseppe and Sanfilippo, Francesco and Simula, Silvano and Lubicz, Vittorio",
    collaboration = "Extended Twisted Mass",
    title = "{Inclusive hadronic decay rate of the {\ensuremath{\tau}} lepton from lattice QCD}",
    eprint = "2308.03125",
    archivePrefix = "arXiv",
    primaryClass = "hep-lat",
    doi = "10.1103/PhysRevD.108.074513",
    journal = "Phys. Rev. D",
    volume = "108",
    number = "7",
    pages = "074513",
    year = "2023"
}

@article{Kellermann:2025pzt,
    author = {Kellermann, Ryan and Hu, Zhi and Barone, Alessandro and Elgaziari, Ahmed and Hashimoto, Shoji and Kaneko, Takashi and J{\"u}ttner, Andreas},
    title = "{Inclusive semileptonic decays from lattice QCD: Analysis of systematic effects}",
    eprint = "2504.03358",
    archivePrefix = "arXiv",
    primaryClass = "hep-lat",
    reportNumber = "KEK-CP-0407, CERN-TH-2025-059",
    doi = "10.1103/kltp-1p1f",
    journal = "Phys. Rev. D",
    volume = "112",
    number = "1",
    pages = "014501",
    year = "2025"
}

@article{Aliberti:2025beg,
    author = "Aliberti, R. and others",
    title = "{The anomalous magnetic moment of the muon in the Standard Model: an update}",
    eprint = "2505.21476",
    archivePrefix = "arXiv",
    primaryClass = "hep-ph",
    reportNumber = "CERN-TH-2025-101, FERMILAB-PUB-25-0344-T, INT-PUB-25-015, IPARCOS-UCM-25-029, KEK Preprint 2025-22, LTH 1403, MITP-25-037, UWThPh 2025-15, UWThPh
  2025-15, ZU-TH 37/25, IPARCOS-UCM-25-029",
    doi = "10.1016/j.physrep.2025.08.002",
    journal = "Phys. Rept.",
    volume = "1143",
    pages = "1--158",
    year = "2025"
}

@article{Buividovich:2025hmn,
    author = "Buividovich, P. V. and Hind, B.",
    title = "{Spectral reconstruction based on dimensional reduction in high-temperature gauge theories}",
    eprint = "2512.23560",
    archivePrefix = "arXiv",
    primaryClass = "hep-lat",
    month = "12",
    year = "2025"
}

@article{Tsuji:2026zku,
    author = "Tsuji, Ryutaro and Hashimoto, Shoji",
    title = "{Spectral reconstruction from Euclidean lattice correlators through singular value decomposition}",
    eprint = "2605.15674",
    archivePrefix = "arXiv",
    primaryClass = "hep-lat",
    reportNumber = "KEK-TH-2837",
    month = "5",
    year = "2026"
}
